\documentclass[reprint,aps,prl,superscriptaddress,amsmath,amssymb,floatfix]{revtex4-2}
\usepackage{graphicx}
\usepackage{bm}
\usepackage{xcolor}
\usepackage[normalem]{ulem}
\usepackage[colorlinks=true,linkcolor=blue,citecolor=blue,urlcolor=blue]{hyperref}

\newcommand{\J}{\bm{J}}
\newcommand{\qv}{\hat{\bm{q}}}
\newcommand{\Umat}{\bm{U}}
\newcommand{\Smat}{\bm{\Sigma}}
\newcommand{\Vmat}{\bm{V}}
\newcommand{\Cmat}{\bm{C}}
\newcommand{\rmax}{\rho}
\newcommand{\smax}{\sigma_{\max}}
\newcommand{\avg}[1]{\big\langle #1 \big\rangle}

\begin{document}

\makeatletter
\g@addto@macro\normalsize{%
  \setlength{\abovedisplayskip}{3pt plus 1pt minus 1pt}%
  \setlength{\belowdisplayskip}{3pt plus 1pt minus 1pt}%
  \setlength{\abovedisplayshortskip}{1pt plus 1pt}%
  \setlength{\belowdisplayshortskip}{2pt plus 1pt}}
\makeatother

\title{A New Route to Chaos\\ through the Geometric Composition of Non-Normal Amplification}

\author{D.~Sornette}
\email{dsornette@ethz.ch}
\affiliation{Institute of Risk Analysis, Prediction and Management (Risks-X),
Southern University of Science and Technology, Shenzhen 518055, China}
\author{V.~R.~Saiprasad}
\affiliation{Institute of Risk Analysis, Prediction and Management (Risks-X),
Southern University of Science and Technology, Shenzhen 518055, China}
\author{V.~Troude}
\affiliation{Institute of Risk Analysis, Prediction and Management (Risks-X),
Southern University of Science and Technology, Shenzhen 518055, China}
\date{September 15, 2026}

\begin{abstract}
Chaos emerges when stretching is repeatedly recycled by reinjection. We uncover a new route to chaos in which the decisive variable is the temporal order of non-normal tangent maps: periodic and chaotic states can share essentially the same one-step stretching statistics while their ordered products acquire opposite Lyapunov growth. We introduce the ordered-product growth rate \(h_L\) over \(L\) successive tangent maps, which reveals how states indistinguishable at one step separate under geometric composition and identifies the finite composition scale at which chaos emerges.
We use this mechanism to establish a new form of global chaos control: minute phase actions reorient the successive non-normal amplification directions so that their geometric composition becomes contracting, suppressing chaos at fixed dissipation without reducing local amplification or targeting a preselected orbit.
\end{abstract}

\maketitle

Chaos requires the repeated stretching of nearby trajectories together with their reinjection into a bounded region of phase space. In multidimensional systems, the local stretching itself need not determine whether this recurrent dynamics is periodic or chaotic. When the tangent Jacobians are non-normal, each step possesses preferred amplification directions, and the perturbation produced by one step is generally reoriented relative to those amplified by the next. What then matters is not only the amplification available at each step, but how these directions are geometrically composed along the orbit.

Non-normal amplification is well known to generate transient growth and to underlie subcritical transitions in hydrodynamics and other non-Hermitian or non-normal systems \cite{trefethen1993,schmid2007,linyoung,neubertcaswell,murphymiller,elganainy,troudePRR}. Yet these results concern what a single linearized operator can amplify. They do not determine how the successive amplification directions generated by a nonlinear trajectory combine over many steps. The same collection of local gains can therefore produce very different asymptotic growth when their orientations are visited in a different geometric sequence \cite{furstenberg1963,daubechieslagarias,oseledets,ginelli2007}.

We show that this geometric composition provides a new route to chaos. In a generalised H\'enon map \cite{henon1976} and in the Ikeda map of a driven optical cavity \cite{ikeda1979,ikeda1980}, stable periodic and chaotic attractors can have essentially indistinguishable one-step spectral statistics while their maximal Lyapunov exponents have opposite signs. The transition is carried by the way successive non-normal amplification directions compound or cancel under reinjection, rather than by a substantial change in the local amplification itself. In the generalised H\'enon map, this geometry is generated autonomously by the orbit, without an imposed switching rule \cite{companionPRE}; in the Ikeda map, chaos emerges at fixed round-trip loss, so the transition is driven by the geometric composition of successive tangent maps rather than by changing dissipation.

To expose this mechanism, we introduce the ordered-product growth rate $h_L$ for the geometric composition of $L$ successive tangent maps. At $L=1$, the periodic and chaotic states are essentially indistinguishable in their one-step spectral statistics; over a finite number of composition steps, their $h_L$ values separate and evolve toward Lyapunov growth of opposite sign. The corresponding composition length identifies the scale over which the geometry of successive amplification directions becomes dynamically decisive \cite{daubechieslagarias,furstenberg1963,oseledets}.

This mechanism also suggests a fundamentally different form of chaos control. Rather than weakening local amplification or stabilizing a preselected unstable periodic orbit \cite{ogy,shinbrot1993,boccaletti,roy1992,pyragas,gills1992}, we act on the geometry of successive amplification directions themselves. In the optical cavity, minute phase modulations alter their geometric composition and globally suppress chaos at fixed dissipation, while leaving the available one-step amplification essentially unchanged.

\textit{The growth of a perturbation is decided by the geometric composition of successive amplification directions.} 
For a smooth map $\bm{x}_{n+1}=\bm{F}(\bm{x}_n)$ with Jacobian $\J_n$ along an orbit,
let $\qv_n$ denote the unit vector along an infinitesimal displacement between two nearby
trajectories,
$
\qv_{n+1}=\frac{\J_n\qv_n}{\lVert\J_n\qv_n\rVert},
$
so that the maximal Lyapunov exponent is the realized asymptotic growth rate,
$
\lambda_1=\avg{\ln\lVert\J_n\qv_n\rVert},
$
with $\avg{\cdot}$ denoting the orbit average. At a single step, the largest possible
amplification is $\smax(\J_n)$, while repeated application of that same Jacobian would
produce an asymptotic growth rate set by its spectral radius $\rmax(\J_n)$, the largest
eigenvalue modulus. For a non-normal Jacobian, $\smax(\J_n)>\rmax(\J_n)$ because the
direction of maximal amplification is generally distinct from its eigendirections.
Along a nonlinear orbit, however, the Jacobian changes from step to step. The direction
amplified by $\J_n$ is therefore reoriented before encountering $\J_{n+1}$, whose own
preferred amplification directions are generally different. Consequently, the
long-time growth of a perturbation depends on the geometric composition of these
successive amplification directions: local amplification can either compound or be
partly cancelled according to their relative orientations. Neither $\rmax(\J_n)$ nor
$\smax(\J_n)$ at one step contains this inter-step geometric information.

What is missing from one-step spectral diagnostics is a quantity that follows how
successive amplification directions are geometrically composed over a finite number of
steps. We therefore introduce the ordered-product growth rate over $L$ consecutive
Jacobians,
\begin{equation}
  h_L=\frac{1}{L}\avg{\ln\rmax\!\big(\J_{n+L-1}\cdots\J_n\big)} ,
  \label{eq:hL}
\end{equation}
which interpolates between the one-step spectral rate and the asymptotic Lyapunov
growth. For $L=1$,
$
h_1=\avg{\ln\rmax(\J_n)},
$
so periodic and chaotic states can remain essentially indistinguishable at the level of
their one-step spectra. As $L$ increases, the successive tangent maps are geometrically
composed and $h_L$ approaches $\lambda_1$: the leading eigenvector of the product aligns
with the leading Oseledets direction at the rate
$e^{-L(\lambda_1-\lambda_2)}$, with $\lambda_2$ the subdominant exponent
\cite{oseledets,ginelli2007}. In the six orbits studied below for the two Ikeda cascades and the
generalised-map pair, $h_{1024}$ agrees with the maximal Lyapunov exponent $\lambda_1$ to within
$9.4\times10^{-4}$ \cite{SM}. The important regime is therefore moderate $L$, where
states that are almost indistinguishable at $L=1$ acquire ordered-product growth rates
of opposite sign. For a periodic--chaotic pair with essentially indistinguishable one-step spectral
statistics, we define the separation length $L^*$ as the smallest composition length
at which the $95\%$ confidence intervals of their respective $h_L$ values, estimated
from $32$ contiguous blocks (End Matter), become disjoint and have opposite signs.
Thus, $L^*$
provides a finite composition scale over which the geometry of successive amplification
directions becomes dynamically decisive. Products of matrices of this type are standard
objects, and their extremal growth over all possible orderings is related to the joint
spectral radius \cite{daubechieslagarias}; here, instead, we follow the single geometric
sequence selected autonomously by the dynamics.

The singular value decomposition makes explicit what is contained in one-step
amplification and what appears only through finite-step geometric composition. Writing
$\J_n=\Umat_n\Smat_n\Vmat_n^{\top}$ and
$\Cmat_n=\Vmat_{n+1}^{\top}\Umat_n$, and labelling the $L$ successive steps by
$i=1,\dots,L$, the product becomes
\begin{equation}
  \J_{n+L-1}\cdots\J_n
  =\Umat\,\Smat_L\,\Cmat_{L-1}\Smat_{L-1}\cdots
  \Cmat_1\Smat_1\,\Vmat^{\top},
  \label{eq:sigmac}
\end{equation}
with $\Umat$ and $\Vmat$ the outer frames of the last and first steps. 
The matrices $\Smat_i$ contain the singular gains at each step, with the maximum
one-step amplification given by $\smax(\J_n)$, the largest diagonal entry of
$\Smat_n$. By contrast, the
couplings
$
\Cmat_i=\Vmat_{i+1}^{\top}\Umat_i
$
encode the relative orientation between the output amplification frame of one step and
the input amplification frame of the next
\cite{daubechieslagarias,troudePRR,furstenberg1963,oseledets,ginelli2007}.
This inter-step geometric information is absent from any one-step spectrum. Two orbits
can therefore share essentially the same statistics of the local gains $\Smat_i$ while
their successive frame couplings $\Cmat_i$ differ, so that amplification compounds in
one case and is cancelled in the other. It is this geometric composition of successive
non-normal amplification directions, rather than their one-step strength, that separates
periodic from chaotic growth. We illustrate this mechanism in a generalised H\'enon map
and in the Ikeda map of a driven optical cavity.

\textit{A generalised H\'enon map exhibiting the geometric-composition route to chaos.}
The ordinary H\'enon map
$x_{n+1}=1-ax_n^2+y_n$, $y_{n+1}=bx_n$~\cite{henon1976}
provides a useful constrained reference because, despite its non-normal Jacobians, it
cannot realize the geometric-composition mechanism in its unrestricted form. A single
state-independent change of coordinates symmetrizes every Jacobian along every orbit,
so the spectral radius of any tangent-map product is bounded by the product of the local
spectral radii. Consequently,
$\lambda_1\le h_1$~\cite{SM}: geometric composition may reduce the local spectral
growth, but it cannot generate asymptotic growth beyond that one-step bound.

\begin{table}[!t]
\caption{Stable periodic and chaotic states for three H\'enon-type constructions:
(i) the ordinary H\'enon map, obtained after the rescaling
$y\to\sqrt{b}\,y$ with $\kappa=1$ and $\eta=c=0$;
(ii) the generalised map with $(a,b,c,\kappa)=(1.1,0.3,0.4,2)$;
and (iii) an autonomous alternating-shear variant in which the two linear couplings
alternate between reciprocal shear factors $\kappa$ and $1/\kappa$, with
$\eta=c=0$ and $(a,b)=(1.1,0.3)$.
Each pair of rows compares a stable periodic state with a chaotic state of the same
construction. $\Delta$ denotes the chaotic-minus-periodic change within each pair, and
the ratios use the six-digit values reported in the Supplemental Material~\cite{SM}.
For the alternating-shear case, $\lambda_1$ and
$h_1$ denote the Lyapunov exponent and one-step spectral rate of the
$(x,y)$ tangent block.}
\label{tab:henon}

\centering
\scriptsize
\setlength{\tabcolsep}{1.8pt}
\begin{tabular}{l l l r r c c}
\hline\hline
Map & Parameter & Attractor & $\lambda_1$ & $h_1$ & $\big|\frac{\Delta h_1}{\Delta\lambda_1}\big|$ & $L^*$ \\
\hline
Ordinary & $a=1.0580$ & period 128 & $-0.0089$ & $0.4534$ & & \\
         & $a=1.0585$ & chaotic    & $+0.0049$ & $0.4539$ & $3.5\%$ & 8 \\
Generalised & $\eta=-0.00180$ & period 64 & $-0.0083$ & $0.464121$ & & \\
            & $\eta=-0.00210$ & chaotic   & $+0.0107$ & $0.464183$ & $0.33\%$ & 8 \\
Alternating$^{\dagger}$ & $\kappa=0.944$ & period 32 & $-0.0277$ & $0.4561$ & & \\
 shear      & $\kappa=0.945$ & chaotic   & $+0.0126$ & $0.4567$ & $1.3\%$ & 4 \\
\hline\hline
\end{tabular}
\end{table}

We therefore construct a minimal extension that preserves the H\'enon architecture while
removing this global geometric constraint,
\begin{equation}
  x_{n+1}=1-ax_n^2+\sqrt{b}\,\kappa y_n,\;\;
  y_{n+1}=\tfrac{\sqrt{b}}{\kappa}\,x_n+\eta\,x_ny_n-c\,y_n^2,
  \label{eq:henondef}
\end{equation}
where $a$ and $b$ retain their usual H\'enon meaning. The factor $\kappa$ redistributes
the linear coupling between the two directions, multiplying the $y\!\to\!x$ term and
dividing the $x\!\to\!y$ term. The crucial ingredient is the bilinear term
$\eta x_ny_n$: it makes the condition for symmetrizing the tangent Jacobian depend on
the current state, so that for constant $\kappa$ and $\eta\neq0$, no single
change of coordinates can symmetrize the entire Jacobian sequence~\cite{SM}. The
successive amplification frames are therefore free to undergo a nontrivial geometric
composition along the orbit, and the bound $\lambda_1\le h_1$ no longer holds.
The term $c>0$ keeps the $y$ dynamics bounded. A constant $\kappa$ by itself carries no
new dynamics, since it is removed exactly by the rescaling
$(x,y)\mapsto(x,\kappa y)$ together with $c\mapsto c/\kappa$.

For $(a,b,c,\kappa)=(1.1,0.3,0.4,2)$, changing $\eta$ by only
$3\times10^{-4}$ takes the system from a stable period-$64$ orbit to chaos
(Table~\ref{tab:henon}). Yet the mean one-step spectral rates differ by only
$(6.22\pm0.03)\times10^{-5}$, against a change $0.0190$ in $\lambda_1$.
The distinction appears only after several tangent maps are composed:
the $h_L$'s of the stable periodic orbit and of the chaotic attractor first acquire opposite signs at $L^*=8$~\cite{SM}. The stable
orbit is especially revealing, since
$h_1=+0.4641>0>\lambda_1$: strong one-step amplification is converted into
asymptotic contraction by the geometric composition of successive amplification
directions. The period-$64$ pair gives the closest one-step match among the four
periodic--chaotic boundaries examined, while period $32$ is the shortest orbit in the
same scan for which comparably close one-step statistics can still be obtained
\cite{SM}.

To isolate reorientation alone, we also consider an autonomous alternating-shear
extension in which the H\'enon couplings switch between reciprocal factors
$\kappa$ and $1/\kappa$ while the added degree of freedom remains neutral
\cite{SM}. For $(a,b)=(1.1,0.3)$, changing $\kappa$ from $0.944$ to $0.945$
changes the $(x,y)$ dynamics from a stable period-$32$ orbit to chaos, although the
one-step rate changes by only $1.3\%$ of the change in $\lambda_\perp$; the finite
products separate the two states already at $L^*=4$ (Table~\ref{tab:henon}).

The three H\'enon constructions progressively release the geometric constraints on
successive tangent maps. In the ordinary map, simultaneous symmetrizability enforces
$\lambda_1\le h_1$; alternating reciprocal shears break the common frame by
reorientation alone; the state-dependent bilinear term removes the constraint
altogether. Correspondingly,
$|\Delta h_1/\Delta\lambda_1|$ falls from $3.5\%$ to $1.3\%$ to $0.33\%$,
while the one-step distributions of $\ln\rmax$ and $\ln\smax$ remain closely matched
\cite{SM}. Thus, as the geometric constraint is lifted, Lyapunov growth becomes
increasingly decoupled from one-step spectral amplification.

To isolate the dynamical role of geometric composition, we change only the inter-step
couplings while preserving every individual Jacobian and every one-step statistic
exactly. We store the Jacobians
along a trajectory, divide the sequence into segments of length $m$, and randomly
permute the segments while preserving the order within each segment. In the SVD
representation of Eq.~\eqref{eq:sigmac}, all $\Smat_i$ are unchanged and only the
junction couplings $\Cmat_i$ are altered. For two adjacent segments, 
exchanging their positions leaves the internal geometric
composition within each segment unchanged and replaces only the coupling at the new
junction between them. In the SVD representation, this operation is
\begin{equation}
\begin{aligned}
  &\Smat_{2m}\Cmat_{2m-1}\cdots\Cmat_{m+1}\Smat_{m+1}\;\Cmat_m\;
   \Smat_m\Cmat_{m-1}\cdots\Cmat_1\Smat_1\\
  \longmapsto\;
  &\Smat_m\Cmat_{m-1}\cdots\Cmat_1\Smat_1\;\widetilde{\Cmat}\;
   \Smat_{2m}\Cmat_{2m-1}\cdots\Cmat_{m+1}\Smat_{m+1},
\end{aligned}
\label{eq:exchange}
\end{equation}
with $\widetilde{\Cmat}=\Vmat_1^\top\Umat_{2m}$. Using
$1.024\times10^6$ stored Jacobians and $32$ permutations per segment length,
the rearranged period-$64$ sequence retains a negative Lyapunov exponent only when
segments of at least $m=8$ consecutive Jacobians are preserved.
At $m=8$, the rearranged Jacobian sequence taken from the period-$64$ orbit has
$\lambda_1=-0.0109$, whereas that taken from the chaotic orbit has
$\lambda_1=+0.0109$~\cite{SM}.
Shuffling the Jacobians individually makes both rates positive
\cite{SM,troudeKesten1,troudeKesten2}. The same eight-step scale therefore appears in
both $L^*$ and the reordering experiment, directly linking stability to the finite-step
geometry of the tangent-map sequence.

\begin{figure}[!t]
\centering
\includegraphics[width=\columnwidth]{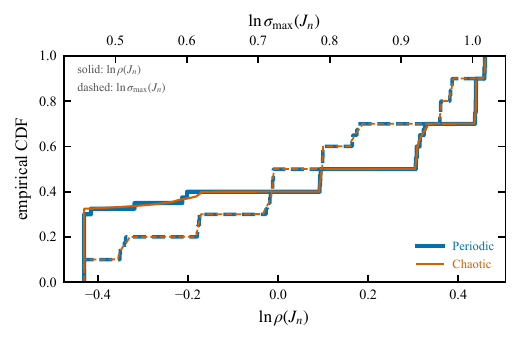}
\caption{Empirical cumulative distributions of $\ln\rmax(\J_n)$ (solid, bottom axis) and
of $\ln\smax(\J_n)$ (dashed, top axis) for the Ikeda map given by Eq.~\eqref{eq:ikeda},
$B=0.65$, $\phi=1.5$, $\beta=3.0846$ (stable period-40 orbit) and $\beta=3.0836$
(chaotic). For both diagnostics, the periodic and chaotic distributions are separated by less
than $1\%$ of their pooled standard deviation in normalized Wasserstein distance.}
\label{fig:cdf}
\end{figure}

\begin{figure}[!t]
\centering
\includegraphics[width=\columnwidth]{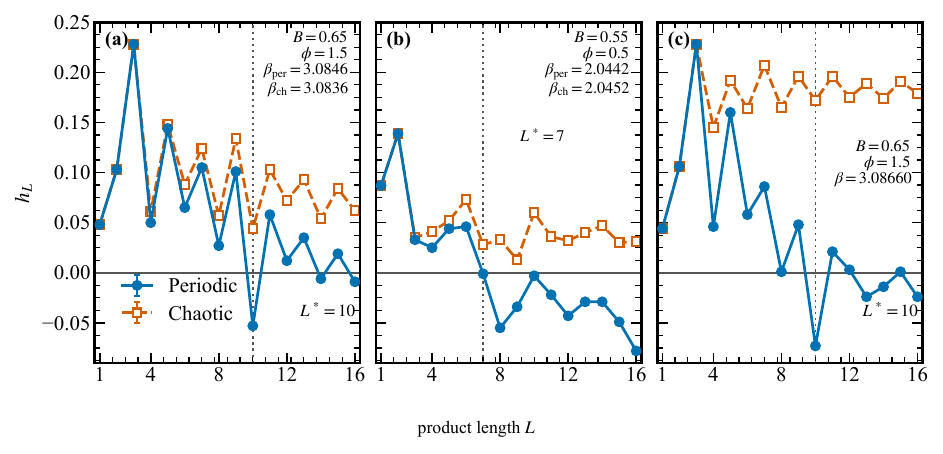}
\caption{Ordered-product growth rate $h_L$ [Eq.~\eqref{eq:hL}] versus the number $L$ of
successively composed tangent maps for three periodic--chaotic Ikeda comparisons:
(a,b) across two period-doubling cascades and (c) for coexisting attractors at
$\beta=3.08660$. Error bars are $95\%$ confidence intervals; vertical dotted lines mark $L^*$.}
\label{fig:hL}
\end{figure}

\textit{Geometric-composition route to chaos in the Ikeda map.}
The Ikeda map describes light circulating in a driven optical cavity with Kerr
nonlinearity~\cite{ikeda1979,ikeda1980}. Writing the complex intracavity field after
the $n$th round trip as $z_n=x_n+i y_n$, and choosing units in which the injected
field is unity,
\begin{equation}
  z_{n+1}=1+Bz_n\exp\!\big[i\big(\phi+\beta|z_n|^2\big)\big],
  \label{eq:ikeda}
\end{equation}
where $B<1$ is the round-trip amplitude transmission, $\phi$ the linear phase, and
$\beta|z_n|^2$ the Kerr nonlinear phase shift. Thus $1-B^2$ is the fractional
intensity loss per round trip.

Because the nonlinear phase depends on $|z_n|^2$, the tangent dynamics is described by
a real $2\times2$ Jacobian acting on $\bm r_n=(x_n,y_n)^\top$. With
$\theta_n=\phi+\beta\lVert\bm r_n\rVert^2$,
\begin{equation}
  \J_n
  =B\,R(\theta_n)\left(\bm I+\bm N_n\right),~
  \bm N_n=2\beta R(\pi/2)\bm r_n\bm r_n^\top ,
  \label{eq:ikedajac}
\end{equation}
where $R(\theta)$ denotes a rotation by $\theta$. Since
$\bm r_n^\top R(\pi/2)\bm r_n=0$, one has $\bm N_n^2=0$.
Thus each tangent step consists of a state-dependent shear, followed by a rotation
and the uniform attenuation $B$. The shear is non-normal and produces amplification
not captured by its eigenvalues. Writing
$s_n=2\beta\lVert\bm r_n\rVert^2$,
$
\smax(\J_n)
=B\,\frac{s_n+\sqrt{s_n^2+4}}{2}
>\rmax(\J_n).
$
Moreover, $\det\J_n=B^2$, so
$\lambda_1+\lambda_2=2\ln B$. At fixed $B$, the transition to chaos therefore
arises from how successive shears and rotations redistribute growth between the two
Lyapunov directions.

For fixed loss, the Ikeda map exhibits periodic--chaotic transitions with almost
unchanged one-step spectral statistics. For $B=0.65$ and $\phi=1.5$,
$\beta=3.0846$ gives a stable period-$40$ orbit with
$\lambda_1=-0.0802$, whereas $\beta=3.0836$ is chaotic with
$\lambda_1=+0.0471$. Yet $h_1$ changes only from $0.047575$ to
$0.047660$, i.e. by $8.5\times10^{-5}$, just $0.07\%$ of the change in
$\lambda_1$, and the full one-step distributions nearly superpose
(Fig.~\ref{fig:cdf})~\cite{SM}. Likewise, for $B=0.55$ and $\phi=0.5$,
the transition from the period-$32$ orbit at $\beta=2.0442$
($\lambda_1=-0.0831$, $h_1=0.084879$) to chaos at $\beta=2.0452$
($\lambda_1=+0.0268$, $h_1=0.084836$) changes $h_1$ by only
$-4.3\times10^{-5}$, or $0.04\%$ of the corresponding change in
$\lambda_1$~\cite{SM}. In contrast, the ordered products separate the two
states at $L^*=10$ and $7$, respectively
[Figs.~\ref{fig:hL}(a),(b)], while segment rearrangement yields opposite
Lyapunov signs at $m=16$ and $4$~\cite{SM}. Both transitions belong to
period-doubling cascades accumulating at
$\beta_\infty\simeq3.08442$ and $2.04481$~\cite{feigenbaum,SM}.

\begin{figure}[!tb]
\centering
\includegraphics[width=\columnwidth]{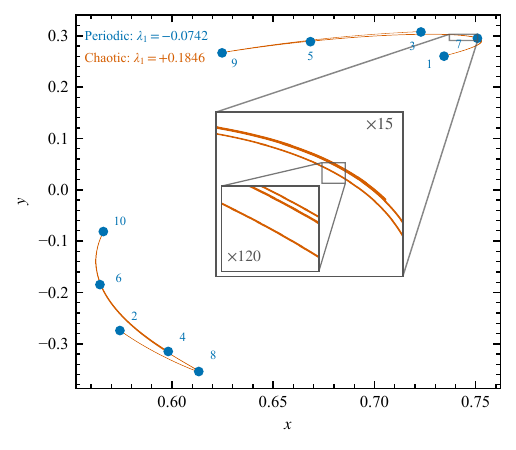}

\caption{Phase portrait of two coexisting attractors of the Ikeda map
[Eq.~\eqref{eq:ikeda}] at identical parameters
$B=0.65$, $\phi=1.5$, and $\beta=3.08660$.
Filled circles show the stable period-$10$ orbit, numbered in visiting order;
thin lines show the chaotic attractor. Successive insets magnify the boxed regions
by factors $15$ and $120$, revealing the splitting of a filament into distinct
bands and the fractal structure of the chaotic attractor ($D\simeq1.18$)~\cite{SM}.}
\label{fig:coexist}
\end{figure}

The same mechanism persists when the competing invariant sets are geometrically
distinct. For $B=0.65$, $\phi=1.5$, and $\beta=3.08660$, just beyond the
boundary crisis at $\beta_c\simeq3.08658$~\cite{grebogi1983,SM}, a stable
period-$10$ orbit with $\lambda_1=-0.0742$ coexists with a chaotic attractor
with $\lambda_1=+0.1846$ (Fig.~\ref{fig:coexist}). Their one-step rates,
$h_1=0.0475$ and $0.0416$, differ by only $0.0059$, or $2.3\%$ of the
Lyapunov-exponent difference, whereas at $L=10$ the ordered-product rates are
$-0.0742$ and $+0.1712$ [Fig.~\ref{fig:hL}(c)]. Thus the separation again
appears through finite-step composition rather than one-step amplification.
This near-invariance is specific to transitions at fixed loss: changing
$B$ from $0.6420$ to $0.6422$ at fixed $\phi=1.5$ and $\beta=3.0836$
changes $h_1$ by $0.0027$, about $6\%$ of the corresponding change in
$\lambda_1$, two orders of magnitude larger than in the two cascades~\cite{SM}.

\textit{Global chaos control through geometric reorientation.}
If chaos is governed by the geometric composition of successive amplification
directions, it should be possible to suppress it by perturbing that composition alone.
We therefore modulate only the round-trip phase of the Ikeda map, leaving both the local dissipation
and the available one-step amplification essentially unchanged, without targeting a
preselected unstable periodic orbit~\cite{ogy,shinbrot1993,boccaletti,roy1992} or
requiring knowledge of a target period~\cite{pyragas}. An intracavity electro-optic
modulator applies a phase shift $u_n$ at each round trip,
\begin{equation}
  z_{n+1}=1+Bz_n\exp\!\big[i\big(\phi+\beta|z_n|^2+u_n\big)\big],~
  |u_n|\le\varepsilon,
  \label{eq:modulation}
\end{equation}
with $\varepsilon=10^{-4}$~rad, four orders of magnitude below the order-one Kerr
phase shift along the orbit. Since $u_n$ changes only the phase,
$\det\J_n=B^2$ and the round-trip loss remain unchanged.

\begin{figure}[!t]
\centering
\includegraphics[width=\columnwidth]{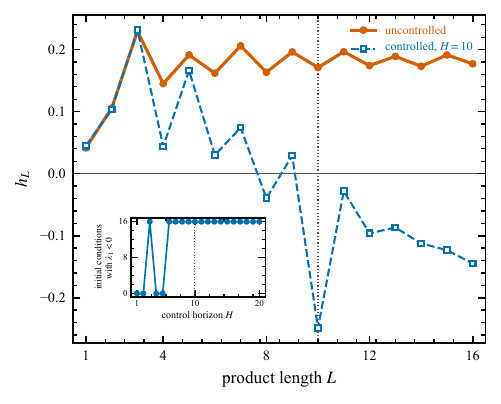}
\caption{Growth rate $h_L$ given by Eq.~\eqref{eq:hL} versus product length $L$ for the chaotic
attractor of Fig.~\ref{fig:coexist}, uncontrolled and controlled dynamics for horizon
$H=10$, the modulation being Eqs.~\eqref{eq:modulation} and \eqref{eq:rule},
$|u_n|\le10^{-4}$~rad; the vertical dotted line marks $L=10$. Inset: number among the $16$ initial
conditions with negative controlled exponent, $H=1$ to $20$.}
\label{fig:control}
\end{figure}

At each step, we choose
$u\in\{-\varepsilon,0,+\varepsilon\}$ to minimize the predicted growth of the current
tangent vector over a horizon $H$,
\begin{equation}
  u_n=\arg\min_{u\in\{-\varepsilon,0,+\varepsilon\}}
  \frac{1}{H}\ln\big\lVert\widehat{\bm{P}}_n^{(H)}(u)\,\qv_n\big\rVert,
  \label{eq:rule}
\end{equation}
where
$\widehat{\bm{P}}_n^{(H)}(u)=\J_{n+H-1}(0)\cdots\J_{n+1}(0)\J_n(u)$
is the predicted $H$-step tangent product with the candidate action applied only at
step $n$~\cite{ushioyamamoto}. We set $H=10$, equal to the separation length
$L^*=10$ of the coexisting pair, without further tuning.

For $\beta=3.08660$, the feedback makes $\lambda_1$ negative for all $16$ tested
initial conditions, which converge to the same period-$70$ closed-loop orbit, not
selected in advance: $\lambda_1$ changes from $+0.1846$ to $-0.3363$.
Crucially, the one-step rate does not decrease, but rises from
$h_1=0.0416$ to $0.0452$, while the ten-step rate reverses from
$h_{10}=+0.1712$ to $-0.2489$ (Fig.~\ref{fig:control}).
The control therefore suppresses chaos by changing the finite-step geometric
composition, not by weakening local amplification.

The control is robust to both the horizon $H$ and errors in the assumed Kerr parameter.
For all $16$ initial conditions, $\lambda_1<0$ for every $H\ge6$, whereas shorter
horizons do not suppress chaos systematically
[Fig.~\ref{fig:control}, inset]~\cite{SM}. Moreover, assuming that the controller computes its
actions using $\beta=3.08660$, while the actual dynamics is generated with $53$
different values of $\beta$ in $[3.08658,3.08710]$, all $848$ controlled trajectories
have $\lambda_1<0$~\cite{SM}. By comparison, a constant phase bias of the same
amplitude succeeds for only $3$ of $25$ benchmark values of $\beta$, versus $23$ for
the feedback rule~\cite{SM}.

\textit{Finite-step geometric composition reveals and controls stability.}
Stability can be tested directly from the Jacobian sequence by examining finite
products of successive tangent maps rather than one-step spectra alone.
The ingredients identified here, non-normal amplification and the geometric
reorientation of successive tangent maps along an orbit, are not restricted to two
dimensions. They provide a general mechanism by which systems with nearly identical
local spectral properties can nevertheless have opposite asymptotic stability.

The same principle also suggests a new route to control. In an optical cavity, a
controller fast enough to act within one round trip could forecast the short tangent
product from a data-driven map and modify only the phase. More generally, whenever
stability cannot be inferred from local spectra, the relevant object is the finite
product that those spectra omit. Our results show that chaos can then be suppressed
by reshaping the geometric composition of successive amplification directions, without
reducing local amplification, changing the dissipation, or selecting a target orbit.
\bibliography{bibliography}

\vspace{3pt}
\noindent\textbf{End Matter}

\textit{Finite-step spectral growth and Lyapunov growth.}
In one dimension the mechanism cannot operate: since
$\rmax(J_n)=|f'(x_n)|$, one has $h_L=\lambda_1$ for every $L$, so identical
one-step statistics imply identical Lyapunov exponents. In higher dimensions, the
$L$-step propagator
$\bm{P}_n^{(L)}=\J_{n+L-1}\cdots\J_n$
defines two distinct growth rates. The finite-time Lyapunov exponent
$
\Lambda_n^{(L)}=\frac{1}{L}\ln\smax\!\big(\bm{P}_n^{(L)}\big)
$
measures the largest amplification over all initial directions, whereas $h_L$
[Eq.~\eqref{eq:hL}] uses the spectral radius of the same product. Hence
$h_L\le\avg{\Lambda_n^{(L)}}$, while both converge to $\lambda_1$ as the product
aligns with the leading Oseledets direction, on a scale set by
$e^{-L(\lambda_1-\lambda_2)}$. For the six Ikeda orbits considered here,
$\lambda_1-\lambda_2=2\lambda_1-2\ln B$ lies between $0.70$ and $1.25$, giving
alignment times much shorter than the measured $L^*$; $L^*$ therefore reflects
finite-step composition rather than directional alignment. On a period-$p$ orbit,
$h_{kp}=\lambda_1$ exactly because the $kp$-step propagator is a power of the
monodromy matrix; for the period-$64$ H\'enon orbit, $h_{64}$ agrees with
$\lambda_1$ within $10^{-4}$~\cite{SM}. For such periodic states, the confidence
interval of $h_L$ reflects only the different starting phases, so $L^*$ is controlled
by the uncertainty of the chaotic member of the pair.

\textit{Stability of the controlled dynamics.}
Because Eq.~\eqref{eq:rule} selects among discrete actions, the closed-loop dynamics is
piecewise defined and has no single Floquet spectrum; stability must therefore be
tested along the realized controlled trajectories. We evolve $160$ independently
controlled pairs of nearby initial conditions for each tested $\beta$. At
$\beta=3.08660$, $141$ of the $160$ pairs have negative fitted separation rates and
$140$ contract below $10^{-10}$; the exceptions cross switching surfaces of the
discrete minimization, showing strong but not uniform contraction~\cite{SM}. This
differs fundamentally from control of a preselected unstable periodic orbit
\cite{ogy,gills1992}: there the trajectory is held near a target unstable orbit whose tangent
product remains expanding, whereas Eq.~\eqref{eq:rule} makes the realized tangent
product itself contracting~\cite{SM}. The comparison uses the same map, phase bound,
initial conditions, and run length, with the target orbit chosen before examining the
closed-loop results~\cite{SM}.

\textit{Numerical and statistical validation.}
Each trajectory discards $10^5$ iterations and retains
$1.024\times10^6$ Jacobians. Uncertainties are $95\%$ confidence intervals from
$32$ contiguous blocks of $3.2\times10^4$ starting positions, avoiding the assumption
that successive iterates are independent. Analytic Jacobians agree with central finite
differences to relative errors below $4.6\times10^{-9}$, and for the Ikeda map
$\lambda_1+\lambda_2=2\ln B$ is recovered to $2.5\times10^{-12}$~\cite{SM}.
The reported attractors are reached from all $16$ initial conditions chosen in their
basins; for the coexisting case, the period-$10$ attractor is reached from $101$ of
$3721$ points on a $61\times61$ grid, confirming a basin of nonzero measure~\cite{SM}.
The same $16$ initial conditions are used throughout the control tests and the
$53$-value $\beta$ sweep. For the chaotic baseline at $\beta=3.08660$, $32$
non-overlapping windows give
$\lambda_1=0.18464$, $h_1=0.04158$, and $h_{10}=0.17120$, with standard deviations
$2.1\times10^{-4}$, $6.3\times10^{-5}$, and $1.5\times10^{-4}$, respectively
\cite{SM}.

\clearpage
\onecolumngrid
\newpage
\begin{center}
{\large\textbf{Supplemental Material for ``A New Route to Chaos through the Geometric Composition of Non-Normal Amplification''}}\\[4pt]
D.~Sornette, V.~R.~Saiprasad, and V.~Troude
\end{center}
\setcounter{secnumdepth}{2}
\setcounter{section}{0}
\setcounter{equation}{0}
\setcounter{figure}{0}
\setcounter{table}{0}
\renewcommand{\thesection}{S\arabic{section}}
\renewcommand{\theequation}{S\arabic{equation}}
\renewcommand{\thefigure}{S\arabic{figure}}
\renewcommand{\thetable}{S\arabic{table}}

This Supplemental Material is self-contained: every symbol used here is defined here, and
cross-references to the Letter are for orientation only.
Section~\ref{sm:sec:methods} fixes estimators, sampling and validation;
Section~\ref{sm:sec:henon} defines the generalised H\'enon map and its diagnostics;
Sections~\ref{sm:sec:henonroute} and \ref{sm:sec:ikedaroute} classify the two period-doubling
transitions by continuation of the periodic branches; Section~\ref{sm:sec:crisis} treats the
crisis at fixed cavity loss, Section~\ref{sm:sec:ordering} presents the reordering tests, and
Section~\ref{sm:sec:control} describes the feedback rule acting on the tangent ordering.

\section{Numerical methods and validation}
\label{sm:sec:methods}

\subsection{Estimators}

For a smooth map $\bm{x}_{n+1}=\bm{F}(\bm{x}_n)$ on $\mathbb{R}^d$, $\J_n=D\bm{F}(\bm{x}_n)$ is
the Jacobian at the $n$th point of an orbit, $\rmax(\J_n)$ its spectral radius (largest
eigenvalue modulus) and $\smax(\J_n)$ its largest singular value (square root of the largest
eigenvalue of $\J_n^{\top}\J_n$). All definitions of this section hold in any finite dimension
$d$. The maps measured in this work act on $\mathbb{R}^2$ (the generalised H\'enon map), on
$\mathbb{R}^3$ (its alternating-shear extension), or on the complex plane (the Ikeda map, written in the
single complex variable $z_n=x_n+iy_n$): because the Ikeda map depends on $|z_n|^2$, it is not
holomorphic, and its tangent dynamics is that of the real $2\times2$ Jacobian on
$\mathbb{C}\simeq\mathbb{R}^2$. The normalized tangent vector $\qv_n$ and the maximal Lyapunov
exponent $\lambda_1$ are
\begin{equation}
  \qv_{n+1}=\frac{\J_n\qv_n}{\lVert\J_n\qv_n\rVert},
  \qquad
  \lambda_1=\lim_{N\to\infty}\frac{1}{N}\sum_{n=0}^{N-1}\ln\lVert\J_n\qv_n\rVert ,
  \label{sm:eq:lyap}
\end{equation}
with $\lambda_2$ the subdominant exponent and $\avg{\cdot}$ the average along the retained orbit.
The ordered $L$-step product and its spectral rate are
\begin{equation}
  \bm{P}_n^{(L)}=\J_{n+L-1}\cdots\J_n ,
  \qquad
  h_L=\frac{1}{L}\avg{\ln\rmax\big(\bm{P}_n^{(L)}\big)} ,
  \label{sm:eq:hL}
\end{equation}
so $h_1=\avg{\ln\rmax(\J_n)}$ is the mean one-step spectral rate and $h_L\to\lambda_1$ as
$L\to\infty$ for a non-degenerate top exponent. A state is called a stable periodic orbit when the
computed $\lambda_1$ is negative and the orbit periodic, chaotic when $\lambda_1$ is positive.

\subsection{Sampling, confidence intervals, separation lengths}

Every production trajectory discards $10^5$ iterations of transient and retains
$1.024\times10^6$ consecutive Jacobians. Uncertainties are $95\%$ confidence intervals built from
the dispersion of block means over $32$ contiguous blocks of $3.2\times10^4$ starting positions,
so successive iterations are never treated as independent samples. On an orbit of minimal period
$p$, the averages in Eqs.~\eqref{sm:eq:lyap} and \eqref{sm:eq:hL} run over the $p$ starting phases
and are exact. Two orbits are told apart by the \emph{separation length $L^*$}, the smallest
tested $L$ at which the two $95\%$ intervals on $h_L$, one per member of a periodic-chaotic pair,
are disjoint and of opposite sign.
Since $h_L$ depends on $L$ in an oscillatory, system-specific way, such a length is an upper bound
at the stated sample size, not a universal threshold, and separation need not persist at every
larger $L$.

\subsection{Validation}

Analytic Jacobians agree with central finite differences to maximum relative errors of
$1.371\times10^{-9}$ for the generalised H\'enon map and $4.535\times10^{-9}$ for the
Ikeda map. For the Ikeda map, defined in Section~\ref{sm:sec:ikedaroute},
the one-step determinant is exactly $\det\J_n=B^2$, with $B$ the round-trip amplitude
transmission, so $\lambda_1+\lambda_2=2\ln B$; this identity is reproduced to
$2.5\times10^{-12}$. On a period-$p$ orbit, $\bm{P}_n^{(kp)}$ is the $k$th power of the monodromy
matrix, hence $h_{kp}=\lambda_1$ exactly whatever the degeneracy of the Floquet multipliers. This
exactness argument does not cover every entry of the tables below, since $1024$ is not a multiple
of every period involved; empirically, over the six sequences studied here the largest measured
deviation is $|h_{1024}-\lambda_1|=9.38\times10^{-4}$. This is an observation for these sequences
and not a theorem for arbitrary matrix products. In the basin audits, all $16$ bounded initial
conditions for each of the six states recover the reported stable periodic orbit or chaotic
orbit, with no failures or escapes.

Two exact null cases provide direct checks of the implementation of
Eq.~\eqref{sm:eq:hL}, and both are limits
in which the ordering has no effect. (i) Isotropic gain: if $\J_n=g_n\bm{O}_n$ with $g_n>0$
and $\bm{O}_n$ orthogonal, then $\bm{P}_n^{(L)}=\big(\prod_k g_{n+k}\big)\bm{O}$ with $\bm{O}$
orthogonal, all of whose eigenvalues have unit modulus, so $h_L=\avg{\ln g}$ for every $L$.
(ii) Frozen tangent map: if $\J_n=\J$ for all $n$, then $\rmax(\J^L)=\rmax(\J)^L$ and
$h_L=\ln\rmax(\J)$ identically. Anisotropic gain and nontrivial inter-step coupling, that is,
non-normal and non-commuting one-step maps, are therefore
both necessary for any dependence of $h_L$ on $L$.

\subsection{Distances between one-step distributions}
\label{sm:subsec:distances}

With $F$ and $G$ the empirical cumulative distribution functions of the same one-step diagnostic
($\ln\rmax$ or $\ln\smax$) on the stable periodic orbit and on the chaotic state,
\begin{equation}
\begin{aligned}
  D_{\rm KS}&=\sup_{t}\big|F(t)-G(t)\big| ,\qquad
  D_2=\Big[\frac{1}{\sigma_p}\int\big(F(t)-G(t)\big)^2dt\Big]^{1/2} ,\\
  D_1&=W_1=\int\big|F(t)-G(t)\big|\,dt=\int_0^1\big|F^{-1}(u)-G^{-1}(u)\big|\,du ,
\end{aligned}
  \label{sm:eq:distances}
\end{equation}
where the second form of $W_1$ is its quantile representation, valid in one dimension; $W_1$ is
reported divided by the pooled standard deviation $\sigma_p$ of the two samples, so that
$W_1/\sigma_p$ is a displacement measured in units of the spread of the diagnostic itself, and
$D_2$ is normalized by $\sigma_p$ for the same reason. 

The difficulty of the Kolmogorov-Smirnov supremum distance $D_{\rm KS}$
for discontinuous distributions is known since the adapted goodness-of-fit test of Conover
(J. Am. Stat. Assoc. \textbf{67}, 591 (1972)); the present question is not the null distribution
but whether $D_{\rm KS}$ is a meaningful distance.
Since $D_{\rm KS}$ is a supremum norm, it records the largest local vertical separation between $F$ and
$G$, irrespective of the width of the region over which that separation occurs, so it need not
shrink as the two distributions become close everywhere else. For the discrete, step-like
cumulative distribution function of a periodic orbit of minimal period $p$, a single
probability-mass jump of size $1/p$ that is displaced horizontally by an arbitrarily small amount
against the continuous chaotic law still contributes a finite $D_{\rm KS}$ of order $1/p$, even
where the two distributions are otherwise close over the rest of their support. 
The integrated distance $D_1=\int|F(t)-G(t)|\,dt$ coincides in one dimension with the
first Wasserstein distance $W_1$ and measures the total discrepancy between $F$ and $G$; the
Cram\'er-von Mises-type distance $D_2=[\int(F(t)-G(t))^2dt]^{1/2}$ measures its typical size.

\begin{figure}[t]
\centering
\includegraphics[width=0.86\textwidth]{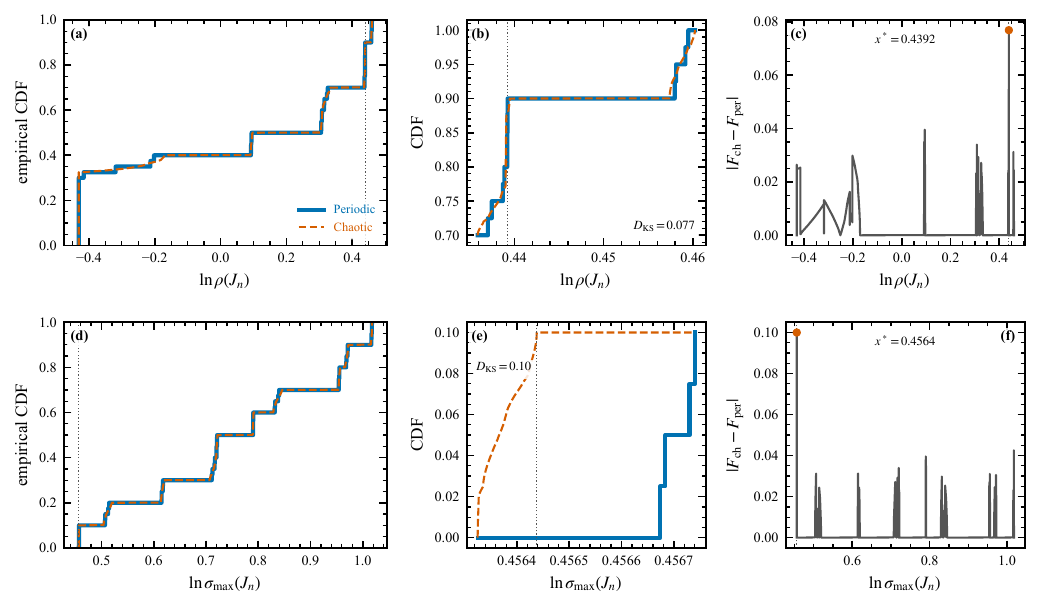}
\caption{Localized origin of the Kolmogorov-Smirnov distance for the Ikeda pair with $B=0.65$
and $\phi=1.5$ of Table~\ref{sm:tab:pairs}. Top row, the $\ln\rmax(\J_n)$ diagnostic; bottom
row, the $\ln\smax(\J_n)$ diagnostic. (a),(d) full empirical cumulative distribution functions
$F_{\rm per}$ (stable period-$40$ orbit) and $F_{\rm ch}$ (chaotic state). (b),(e) the same
functions enlarged around the maximizing point, $x^*=0.439202$ with
$F_{\rm per}(x^*)=0.80$ and $F_{\rm ch}(x^*)=0.877$ for $\ln\rmax$, and
$x^*=0.456438$ with $F_{\rm per}(x^*)=0$ and $F_{\rm ch}(x^*)=0.10$ for $\ln\smax$.
(c),(f) the pointwise difference $|F_{\rm ch}(x)-F_{\rm per}(x)|$, peaking at
$D_{\rm KS}=0.077$ and $0.10$ respectively while remaining small over most of the
support.}
\label{sm:fig:ksvisual}
\end{figure}

A schematic example makes the mechanism explicit. Let $F_1(x)$ equal $0.8$ for $x<x_0$ and
$0.9$ for $x\ge x_0$, and let $F_2$ be the same function displaced to $x_0+\epsilon$. For every
$\epsilon>0$, however small, $\sup_x|F_1(x)-F_2(x)|=0.1$, whereas
$W_1=\int|F_1-F_2|\,dx=0.1\,\epsilon$ vanishes with the displacement. The supremum distance
therefore does not go to zero as the two distributions become identical up to a horizontal
shift, while the integrated distance does, which is the notion of closeness relevant here. For a stable orbit of minimal period $p$ each distinct value of the diagnostic
has probability $m/p$, with $m$ its multiplicity, so a displacement of one jump against the
continuous chaotic law produces a Kolmogorov-Smirnov distance of that order. The period-40 orbit
of the $B=0.65$ pair realizes only ten distinct values of $\ln\smax$, each of weight $1/10$,
which is the origin of the value $D_{\rm KS}=0.10$ of Fig.~\ref{sm:fig:ksvisual}(e); the
value $0.077$ for $\ln\rmax$ is the offset of the chaotic law at one such jump, the scale
$1/10$ being an upper bound. The Kolmogorov-Smirnov distance is not
wrong: it states that at one threshold the proportions on the two sides differ by about
$7.7\%$ or $10\%$. It must not be read as ``the distributions differ by $0.1$''. 

$D_{\rm KS}$, $D_1$ and $D_2$ therefore answer three different questions, the largest local
discrepancy, the total discrepancy and the typical distributed discrepancy, and a pair of
distributions can give a large $D_{\rm KS}$ together with small $D_1$ and $D_2$ when the
discrepancy occupies a narrow region of the support.

For the Ikeda pair of $B=0.65$, $\phi=1.5$ compared in Table~\ref{sm:tab:pairs}, the maximizing
point of the $\ln\rmax(\J_n)$ distributions is $x^*=0.439202$, at which $F_{\rm per}(x^*)=0.80$
and $F_{\rm ch}(x^*)=0.877$, so $D_{\rm KS}=|F_{\rm ch}(x^*)-F_{\rm per}(x^*)|=0.077$: 
a direct instance of the localized-jump mechanism above, since the
periodic orbit's cumulative distribution function crosses the value $0.8$ in one discrete step and
the chaotic law's crossing of the same value falls close by but not exactly there.

Under one common numerical protocol, the three periodic-chaotic comparisons of
Table~\ref{sm:tab:pairs} give the distances of Table~\ref{sm:tab:distances}. For the $\ln\smax$ diagnostic of the Ikeda pair with
$B=0.65$, the value $D_{\rm KS}=0.10$ is attained at the left edge of the periodic
support, at $x^*=0.456438$, where $F_{\rm per}=0$ while $F_{\rm ch}=0.10$: this is the
localized-jump mechanism above in its extreme form, visible in
Fig.~\ref{sm:fig:ksvisual}(e),(f). Under the same protocol, the pair with $B=0.55$ gives
$D_{\rm KS}=0.0564$ for both diagnostics.

\begin{table}[t]
\caption{Distances of Eq.~\eqref{sm:eq:distances} between the one-step distributions of a
stable periodic orbit and of its chaotic partner, for the three comparisons of
Table~\ref{sm:tab:pairs} and for each of the two one-step diagnostics, all evaluated under
one common numerical protocol, together with the ordinary H\'enon pair of
Section~\ref{sm:sec:henon} for comparison. $D_{\rm KS}$ is the supremum distance,
$W_1/\sigma_p$ the first Wasserstein distance divided by the pooled standard deviation,
and $D_2$ the normalized quadratic distance. Dashes mark a distance not measured for that
pair.}
\label{sm:tab:distances}
\centering
\begin{tabular}{l l c c c}
\hline\hline
Comparison & Diagnostic & $D_{\rm KS}$ & $W_1/\sigma_p$ & $D_2$\\
\hline
Generalised H\'enon         & $\ln\rmax(\J_n)$ & $0.0348$ & $0.00199$ & $0.00377$\\
                            & $\ln\smax(\J_n)$ & $0.0264$ & $0.00147$ & $0.00310$\\
Ikeda, $B=0.65$, $\phi=1.5$ & $\ln\rmax(\J_n)$ & $0.0769$ & $0.00786$ & $0.01124$\\
                            & $\ln\smax(\J_n)$ & $0.1000$ & $0.00455$ & $0.00970$\\
Ikeda, $B=0.55$, $\phi=0.5$ & $\ln\rmax(\J_n)$ & $0.0564$ & $0.00924$ & $0.01186$\\
                            & $\ln\smax(\J_n)$ & $0.0564$ & $0.00424$ & $0.00821$\\
Alternating shear           & $\ln\rmax(\J_n)$ & $0.0367$ & $0.00518$ & $0.00887$\\
                            & $\ln\smax(\J_n)$ & $0.0363$ & $0.00501$ & $0.00870$\\
Ordinary H\'enon, $b=0.3$   & $\ln\rmax(\J_n)$ & $0.0250$ & $0.00542$ & --\\
                            & $\ln\smax(\J_n)$ & $0.0250$ & $0.00401$ & --\\
\hline\hline
\end{tabular}
\end{table}

\section{The generalised H\'enon map and its diagnostics}
\label{sm:sec:henon}

\subsection{Definition}

The generalised H\'enon map is
\begin{align}
  x_{n+1}&=1-a\,x_n^{2}+\sqrt{b}\,\kappa\,y_n ,\nonumber\\
  y_{n+1}&=\frac{\sqrt{b}}{\kappa}\,x_n+\eta\,x_ny_n-c\,y_n^{2} ,
  \label{sm:eq:henondef}
\end{align}
with $a$ and $b>0$ the H\'enon parameters, $\kappa>0$ a constant shear redistributing the linear
coupling between the two directions, $c$ the coefficient of the quadratic term that bounds the
$y$ dynamics, and $\eta$ the amplitude of the cross-term. All values below use
$(a,b,c,\kappa)=(1.1,0.3,0.4,2)$, only $\eta$ being varied. The Jacobian is
\begin{equation}
  \J_n=
  \begin{pmatrix}
    -2a x_n & \sqrt{b}\,\kappa\\[1.5mm]
    \dfrac{\sqrt{b}}{\kappa}+\eta\,y_n & \eta\,x_n-2c\,y_n
  \end{pmatrix} ,
  \label{sm:eq:henonjac}
\end{equation}
so $\eta$ acts both on the orbit, through $\eta x_ny_n$ in Eq.~\eqref{sm:eq:henondef}, and on the
tangent dynamics, through the entries $\eta y_n$ and $\eta x_n$ of Eq.~\eqref{sm:eq:henonjac}.

\subsection{Exact statements about symmetrizability}

\emph{(a) The ordinary H\'enon family is simultaneously symmetrizable.} For
$x_{n+1}=1-ax_n^2+y_n$, $y_{n+1}=bx_n$ with $b>0$, the Jacobian has entries $(-2ax_n,1)$ on
its first row and $(b,0)$ on its second, and the fixed, state-independent diagonal change of
coordinates $\bm{T}={\rm diag}(\sqrt{b},1)$ gives
$\bm{T}\J_n\bm{T}^{-1}=\left(\begin{smallmatrix}-2ax_n&\sqrt{b}\\ \sqrt{b}&0\end{smallmatrix}\right)$,
symmetric at every point of every orbit.

\emph{(b) The constant shear is removable at every $\eta$.} The
change of variable $Y_n=\kappa y_n$ in Eq.~\eqref{sm:eq:henondef} gives
$x_{n+1}=1-ax_n^2+\sqrt{b}\,Y_n$,
$Y_{n+1}=\sqrt{b}\,x_n+\eta\,x_nY_n-(c/\kappa)Y_n^2$, that is the same family with
$\kappa=1$ and
$c\to c/\kappa$, for every value of $\eta$. A fixed change of coordinates removes
$\kappa$ entirely, so no dynamical
invariant can depend on $\kappa$ alone and the shear amplitude is never the control
parameter. What is special for $\eta=0$ is symmetry: there the fixed transformation
$\bm{T}={\rm diag}(1/\kappa,1)$ turns both off-diagonal entries of Eq.~\eqref{sm:eq:henonjac}
into $\sqrt{b}$ and symmetrizes every Jacobian. This exact removability was checked numerically
for $(\kappa,c)=(2,0.4)$ and its conjugate $(1,0.2)$: the mean one-step spectral rate agrees to
eight digits, $h_1=0.46373747$ for both, while the finite-trajectory estimates of the maximal
Lyapunov exponent, $\lambda_1=-0.01794305$ against $-0.01794267$, differ only in the seventh
digit, a numerical rather than physical discrepancy, since the coordinate conjugacy removes the
constant shear exactly.

\emph{Consequence.} Let $\bm{T}$ be fixed and invertible with $\bm{S}_n=\bm{T}\J_n\bm{T}^{-1}$
symmetric for all $n$. Then $\lVert\bm{S}_n\rVert_2=\rmax(\bm{S}_n)=\rmax(\J_n)$, and
submultiplicativity of the spectral norm applied to
$\bm{P}_n^{(N)}=\bm{T}^{-1}\bm{S}_{n+N-1}\cdots\bm{S}_n\bm{T}$ gives
\begin{equation}
  \frac{1}{N}\ln\big\lVert\bm{P}_n^{(N)}\big\rVert_2
  \le\frac{\ln\!\big(\lVert\bm{T}\rVert_2\lVert\bm{T}^{-1}\rVert_2\big)}{N}
  +\frac{1}{N}\sum_{k=0}^{N-1}\ln\rmax(\J_{n+k})
  \ \xrightarrow[N\to\infty]{}\ h_1 ,
\end{equation}
hence $\lambda_1\le h_1$. In a simultaneously symmetrizable family, the asymptotic growth rate can
never exceed the mean one-step spectral rate: ordering cannot compound a locally subcritical
spectrum into growth, so $h_1<0<\lambda_1$ is structurally forbidden, while $h_1>0>\lambda_1$, in
which ordering instead compounds a locally supercritical spectrum into decay, remains permitted
and is the case measured here.

\emph{(c) The cross-term breaks simultaneous symmetrizability.} A fixed invertible
$\bm{T}$ with $\bm{T}\J_n\bm{T}^{-1}$ symmetric for every Jacobian of the family exists
if and only if some symmetric positive-definite $\bm{G}=\bm{T}^{\top}\bm{T}$ satisfies
$\bm{G}\J_n=\J_n^{\top}\bm{G}$ for all $n$. For a $2\times2$ Jacobian with entries
$(p,q;r,s)$ this is a single linear condition on $\bm{G}$,
$q\,g_{11}+(s-p)\,g_{12}-r\,g_{22}=0$. For $\eta=0$, the conditions of all Jacobians share
the one-dimensional kernel $\bm{G}\propto{\rm diag}(1/\kappa^{2},1)$, which is the
transformation of (b). At $\eta\neq0$ the coefficient vectors $(q,\,s-p,\,-r)$ vary with
the state and span all of $\mathbb{R}^{3}$ along the orbit, so no fixed symmetrizer
exists.
Each individual Jacobian with $q\,r>0$ remains symmetrizable in isolation, so the
obstruction is collective, set by the orbit as a whole; the bound $\lambda_1\le h_1$ no longer
follows from the structure of the family. The deformation
is smooth, autonomous and two-dimensional: no third variable, no switching rule, no modulo operation.

\subsection{Measured pair and one-step distributions}

For $(a,b,c,\kappa)=(1.1,0.3,0.4,2)$ the two attractors of the measured pair have $\eta$ values
that differ by $3\times10^{-4}$ (Table~\ref{sm:tab:pairs}, first block). Their mean one-step
spectral rates agree to $\Delta h_1=6.21757\times10^{-5}$, with $95\%$ confidence interval
$[6.18349,6.25164]\times10^{-5}$ from $32$ blocks, while $\lambda_1$
changes sign, so
$|\Delta h_1/\Delta\lambda_1|=6.21757\times10^{-5}/1.899\times10^{-2}=3.27\times10^{-3}$. The
stable periodic member has one-step rate $h_1=0.464121$, positive together with $\lambda_1<0$, and
the chaotic member realizes as $\lambda_1$ only $0.010657/0.464183=2.3\%$ of its own
one-step rate.

A period-$4$ orbit of the same map and its chaotic twin were tested as a shorter alternative
to the period-$64$ pair above. Their one-step distributions no longer stay close: for
$\ln\rmax$, $D_{\rm KS}=0.25$ and $W_1/\sigma_p=0.126$, an order of magnitude larger than the
period-$64$ pair's one-step distances. The period-$4$ pair is therefore not used, and the
period-$64$ generalised-H\'enon pair of Table~\ref{sm:tab:pairs} is retained.

\begin{figure}[t]
\centering
\includegraphics[width=0.7\textwidth]{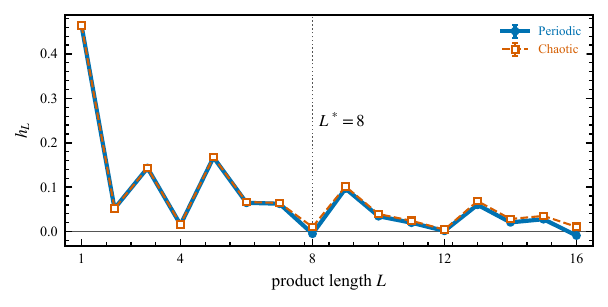}
\caption{Ordered-product rate $h_L$ of Eq.~\eqref{sm:eq:hL} versus product length $L$ for the
generalised H\'enon pair of Table~\ref{sm:tab:pairs}, $(a,b,c,\kappa)=(1.1,0.3,0.4,2)$: stable
period-$64$ orbit for $\eta=-0.00180$ and chaotic state for $\eta=-0.00210$, with $95\%$
confidence intervals. The two mean rates first take opposite sign at composition length
$L^*=8$, marked by the dotted line.}
\label{sm:fig:henongeneralisedhL}
\end{figure}

Across the three periodic-chaotic pairs of Table~\ref{sm:tab:pairs}, the normalized Wasserstein
distances of Eq.~\eqref{sm:eq:distances} between the one-step distributions of $\ln\rmax$ and of
$\ln\smax$ lie between $1.47\times10^{-3}$ and $9.24\times10^{-3}$, below one percent of the
pooled standard deviation in every case, while the Kolmogorov-Smirnov distances reach $D_{\rm KS}=0.10$
(Table~\ref{sm:tab:distances}). The
two statements are consistent: a periodic state of minimal period $p$ has an atomic one-step law
supported on $p$ values, whereas its chaotic partner has a continuous law organized into the same
narrow bands, so the two cumulative distribution functions stay close everywhere but cross in a
staircase pattern, which a supremum norm registers and an integrated distance does not.

\begin{table}[t]
\caption{Measured periodic-chaotic pairs. Top block: generalised H\'enon map of
Eq.~\eqref{sm:eq:henondef} for $(a,b,c,\kappa)=(1.1,0.3,0.4,2)$, control parameter the cross-term
amplitude $\eta$. Lower blocks: Ikeda map of Eq.~\eqref{sm:eq:ikeda} at round-trip transmission
$B$ and linear phase $\phi$, control parameter the Kerr coefficient $\beta$.
$h_1=\avg{\ln\rmax(\J_n)}$ is the mean one-step spectral rate, $\lambda_1$ the maximal Lyapunov
exponent of Eq.~\eqref{sm:eq:lyap} and $L^*$ the separation length defined in
Section~\ref{sm:sec:methods}. Chaotic entries use $1.024\times10^6$ retained Jacobians after a
$10^5$-iterate transient; periodic entries follow the convention of
Section~\ref{sm:sec:methods}, in which the averages run over the $p$ starting phases of the
orbit.}
\label{sm:tab:pairs}
\centering
\begin{tabular}{l l l c c c}
\hline\hline
System & Control value & Attractor & $h_1$ & $\lambda_1$ & $L^*$\\
\hline
Generalised H\'enon & $\eta=-0.00180$ & period 64 & $+0.464121$ & $-0.008330$ & \\
                 & $\eta=-0.00210$ & chaotic   & $+0.464183$ & $+0.010657$ & $8$\\
\hline
Ikeda, $B=0.65$, $\phi=1.5$ & $\beta=3.0846$ & period 40 & $+0.047575$ & $-0.080239$ & \\
                            & $\beta=3.0836$ & chaotic   & $+0.047660$ & $+0.047105$ & $10$\\
\hline
Ikeda, $B=0.55$, $\phi=0.5$ & $\beta=2.0442$ & period 32 & $+0.084879$ & $-0.083108$ & \\
                            & $\beta=2.0452$ & chaotic   & $+0.084836$ & $+0.026835$ & $7$\\
\hline\hline
\end{tabular}
\end{table}

\subsection{Four boundaries of one parameter scan, and the ordinary H\'enon map}

The near-invariance of the one-step statistics is not a property of one bracketed boundary. Four
distinct periodic-chaotic boundaries of the same $\eta$ scan give the values of
Table~\ref{sm:tab:fourboundaries}: $\Delta\lambda_1$ between $+0.019$ and $+0.034$, one-step
shifts $\Delta h_1$ between $-4.5\times10^{-4}$ and $+7.3\times10^{-5}$ without even a consistent
sign, and per-boundary ratios $|\Delta h_1/\Delta\lambda_1|$ between $0.26\%$ and $1.3\%$. These
are four distinct boundaries encountered along one parameter scan and not statistically
independent replications, and the composition lengths in the last column of
Table~\ref{sm:tab:fourboundaries} use the mean-sign criterion, namely the first tested length $L_{\rm sign}$ at
which the two mean rates $h_L$ have opposite signs, rather than the
confidence-interval criterion of Section~\ref{sm:sec:methods} defining $L^*$. The two criteria do not always
agree: for the Ikeda pair with $B=0.65$ the mean-sign criterion returns $L_{\rm sign}=16$ where the
confidence-interval criterion of Table~\ref{sm:tab:pairs} returns $L^*=10$.

The period of the stable member sets how close the one-step statistics of the two members can
be. With the boundaries of the $\eta$ scan refined to $10^{-7}$, pairs built on the period-32,
period-16 and period-8 windows give one-step distances larger than those of the period-64 pair
by average factors of $1.5$, $4.5$ and $12.8$ over the six distance measures of
Section~\ref{sm:subsec:distances} (worst factors $1.8$, $6.5$ and $22.2$): the period-32 pair
remains comparable to the reference, and the shorter periods do not. Period $32$ is therefore
the smallest period for which the one-step statistics of the two members remain close.

\begin{table}[t]
\caption{Four periodic-chaotic boundaries of the same $\eta$ scan of the generalised H\'enon map
for $(a,b,c,\kappa)=(1.1,0.3,0.4,2)$. Each row gives the cross-term amplitude of the stable
periodic orbit and of the chaotic member, the label of the periodic attractor, and the
chaotic-minus-periodic differences
of the maximal Lyapunov exponent and of the mean one-step spectral rate
$h_1=\avg{\ln\rmax(\J_n)}$. The last column is the first tested composition length $L_{\rm sign}$ at which the
two mean rates $h_L$ have opposite signs. These are distinct boundaries of one scan, not
statistically independent replications.}
\label{sm:tab:fourboundaries}
\centering
\begin{tabular}{l r r r r c}
\hline\hline
Periodic attractor & $\eta_{\rm reg}$ & $\eta_{\rm cha}$ & $\Delta\lambda_1$ & $\Delta h_1$
 & $L_{\rm sign}$\\
\hline
Period 48              & $-0.003145$ & $-0.003135$ & $+0.03446$ & $-4.51\times10^{-4}$ & $16$\\
Narrow periodic window & $-0.002720$ & $-0.002700$ & $+0.02829$ & $-1.80\times10^{-4}$ & $16$\\
Period 96              & $-0.002240$ & $-0.002220$ & $+0.02778$ & $+7.3\times10^{-5}$  & $16$\\
Period 64 (reference)  & $-0.001800$ & $-0.002100$ & $+0.01897$ & $+6.22\times10^{-5}$ & $8$\\
\hline\hline
\end{tabular}
\end{table}

Such a ratio depends on how tightly a boundary is bracketed, which is why it is reported as a
range and not as a figure of merit. The same transition-local analysis applied to the
\emph{ordinary} H\'enon map for $b=0.3$, increasing $a$ from $1.0580$ to $1.0585$, moves the
stable periodic attractor, of period $128$, with $\lambda_1=-0.008860$ and $h_1=0.453410$, to a
chaotic state with $\lambda_1=+0.004943$ and $h_1=0.453896$, giving
$|\Delta h_1/\Delta\lambda_1|=3.52\%$, an order of magnitude larger at comparable bracketing, and
the two mean rates $h_L$ first take opposite sign at composition length $L^*=8$. The one-step distances
between the two states are $D_{\rm KS}=0.0250$ for both $\ln\rmax$ and $\ln\smax$, with normalized
Wasserstein distances $W_1/\sigma_p=0.00542$ ($\ln\rmax$) and $0.00401$ ($\ln\smax$): they are entered in Table~\ref{sm:tab:distances} beside the
other three comparisons, so that the closeness of the one-step distributions can be read
across all four cases at once.
The bracket-independent statement is not the ratio but the continuity of the one-step statistics
across a boundary at which $\lambda_1$ changes sign.

\subsection{Three-dimensional alternating-shear extension}

A third construction realizes ordering-driven growth from a locally subcritical spectrum
inside the H\'enon family. Its purpose is to isolate the reorientation: a constant shear
$\kappa$ is removed by a fixed rescaling and is therefore not a parameter of the dynamics,
whereas two shears applied in turn cannot both be removed by any single rescaling. A third
variable $w_n$ is appended to the two-dimensional state to make the alternation autonomous,
\begin{equation}
  x_{n+1}=1-ax_n^2+\sqrt{b}\,w_n\,y_n,\qquad
  y_{n+1}=\frac{\sqrt{b}}{w_n}\,x_n,\qquad
  w_{n+1}=\frac{1}{w_n},\qquad w_0=\kappa,
  \label{sm:eq:clocked}
\end{equation}
so the shear takes the values $\kappa$ and $1/\kappa$ in turn: the frame rotation imposed
by an explicit switching rule in the engineered constructions of Ref.~\cite{companionPRE} is
here produced by an autonomous third coordinate. Because $w_n$ enters the updates of $x$
and $y$ only through the shear, the $(x,y)$ Jacobian block
$\begin{pmatrix}-2ax_n & \sqrt{b}\,w_n\\ \sqrt{b}/w_n & 0\end{pmatrix}$ has determinant $-b$
and characteristic polynomial $\mu^2+2ax_n\mu-b=0$, free of the shear at every state: the
one-step eigenvalue law is blind to a parameter with which the long-time growth moves.
Unlike the constant shear of Section~\ref{sm:sec:henon}, the alternation cannot be removed
by any single constant rescaling, because the two substeps require different rescalings.
The two-dimensional map of Eq.~\eqref{sm:eq:henondef} is the object of the distribution and
permutation measurements of the Letter; the alternating-shear extension isolates the
reorientation from every other change and is the third case of Table~I of the
Letter.

The update $w_{n+1}=1/w_n$ satisfies $w_{n+2}=w_n$ identically, so its own two-step
multiplier is $1$ and the Lyapunov exponent along the $w$ direction is exactly zero for
every $\kappa$: the third variable never itself contributes to the growth or decay of a
perturbation. Because the $w$ update does not depend on $x$ or $y$, the full
three-dimensional Jacobian is block-triangular, with the $(x,y)$ block driven by the
exogenous, exactly period-two shear sequence $\kappa,1/\kappa,\kappa,\ldots$ and separated
from the $w$ direction. The diagnostics reported here are therefore the transverse exponent
$\lambda_\perp$, computed from this $(x,y)$-block cocycle alone, and the matching transverse
ordered rate $h_L^\perp$ of Eq.~\eqref{sm:eq:hL}, and not a full three-dimensional
$\lambda_1$, which the exactly null $w$ mode would pull toward zero.

\begin{figure}[t]
\centering
\includegraphics[width=\textwidth]{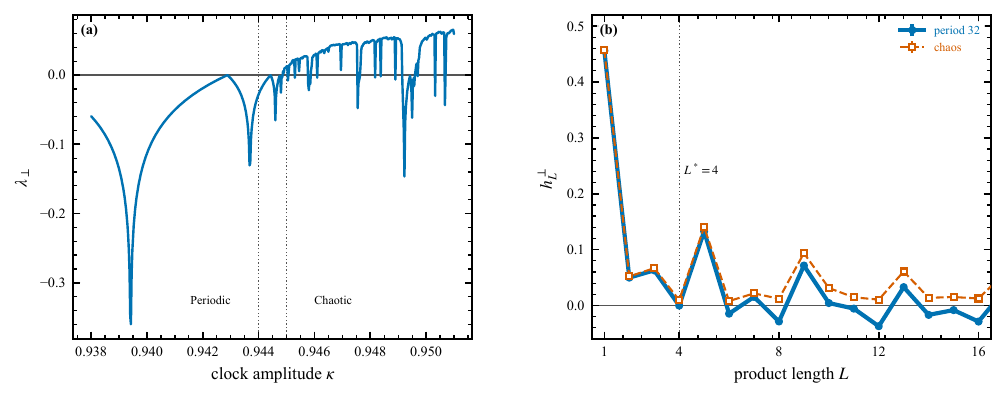}
\caption{Transverse diagnostics of the alternating-shear extension of
Eq.~\eqref{sm:eq:clocked} for $(a,b)=(1.1,0.3)$. (a)~Transverse exponent $\lambda_\perp$
versus the shear amplitude $\kappa$ over $[0.936,0.952]$, with the regular and chaotic
ranges named and vertical dotted lines at the two selected values $\kappa=0.944$
(transversely stable period-$32$ response) and $\kappa=0.945$ (chaotic response).
(b)~Transverse ordered-product rate $h_L^\perp$ of Eq.~\eqref{sm:eq:hL}, computed from the
$(x,y)$-block cocycle, versus product length $L$ for both responses; the dotted line marks
the separation length, $L^*=4$.}
\label{sm:fig:alternatingshear}
\end{figure}

A search across the shear amplitude $\kappa$, at the same $b=0.3$ as the two-dimensional
pair of Table~\ref{sm:tab:pairs} and with $a=1.1$ fixed, gives a transversely stable
response of period $32$ for $\kappa=0.944$, with $\lambda_\perp=-0.027707$ and transverse
one-step rate $h_1^\perp=0.456143$ (the $L=1$ value of $h_L^\perp$), and a chaotic response
for $\kappa=0.945$, with $\lambda_\perp=+0.012643$ and $h_1^\perp=0.456658$
(Fig.~\ref{sm:fig:alternatingshear}). The one-step rate therefore moves by
$|\Delta h_1^\perp/\Delta\lambda_\perp|=5.15\times10^{-4}/4.035\times10^{-2}=1.28\%$ of the
change in the transverse exponent, printed as $1.3\%$ in Table~I of the Letter, and the two transverse ordered rates first take opposite
sign at composition length $L^*=4$. This pair is the one
used for the alternating-shear row of the Letter's Table~I.

The $b=0.3$ pair above is used throughout the alternating-shear comparison, matching the value
of $b$ used elsewhere in the H\'enon discussion.

\section{Route to chaos in the generalised H\'enon map}
\label{sm:sec:henonroute}

\subsection{Scale of the cross-term along the attractor}

\begin{figure}[t]
\centering
\includegraphics[width=0.6\textwidth]{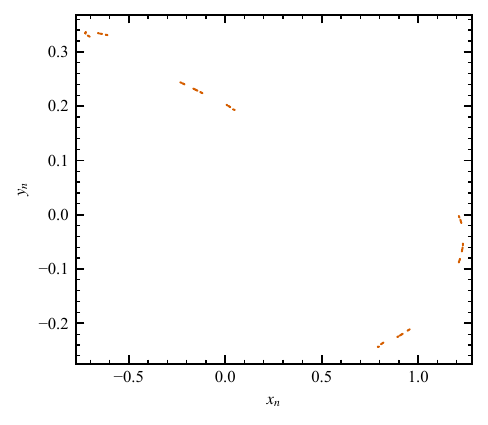}
\caption{Attractor of the generalised H\'enon map of Eq.~\eqref{sm:eq:henondef} for
$(a,b,c,\kappa)=(1.1,0.3,0.4,2)$ and cross-term amplitude $\eta=-0.00210$: $y_n$ versus $x_n$,
plotted after a transient of $10^5$ iterations. The attractor is band chaos of Kaplan-Yorke
dimension $D_{\rm KY}\simeq1.01$: the points spread along thin one-dimensional arcs while
the transverse extent collapses, and no magnification reveals transverse structure, which is
why no zoom is shown. That the chaotic attractor is essentially
one-dimensional supports the statement of the Letter: the transition is not carried by the
local geometry of the set but by the order of the tangent maps along it.}
\label{sm:fig:henonattractor}
\end{figure}

\begin{figure}[t]
\centering
\includegraphics[width=\textwidth]{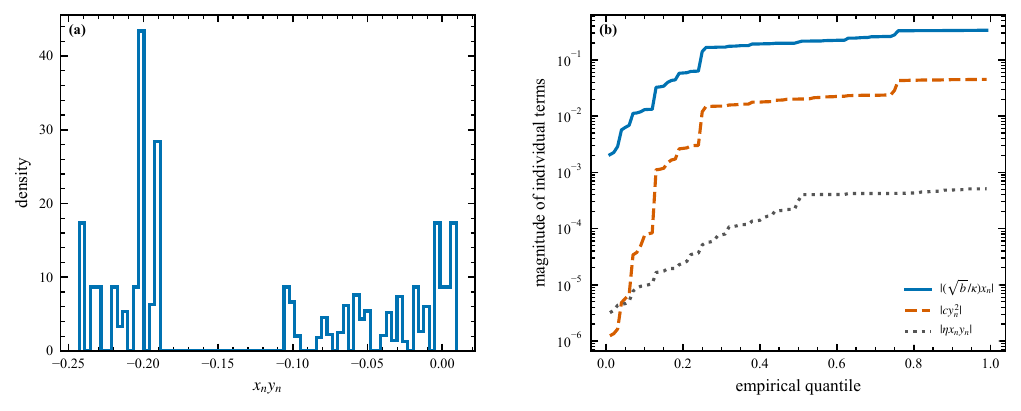}
\caption{Same orbit as Fig.~\ref{sm:fig:henonattractor} ($\eta=-0.00210$,
$(a,b,c,\kappa)=(1.1,0.3,0.4,2)$, $10^5$-iterate transient discarded). (a)~Empirical
distribution of the product $x_ny_n$, the factor multiplying $\eta$ in the $y$ update of
Eq.~\eqref{sm:eq:henondef}. (b)~Magnitudes of the three contributions to $y_{n+1}$, namely
the linear term $(\sqrt{b}/\kappa)x_n$, the quadratic term $c\,y_n^2$ and the cross-term
$\eta\,x_ny_n$, each plotted against its own empirical quantile.}
\label{sm:fig:henonterms}
\end{figure}

For $\eta=-0.00210$, after a transient of $10^5$ steps, the retained chaotic orbit of
Eq.~\eqref{sm:eq:henondef} occupies $-0.72430\lesssim x\lesssim1.23016$,
$-0.24273\lesssim y\lesssim0.33593$ and $-0.24278\lesssim xy\lesssim0.00934$
(Figs.~\ref{sm:fig:henonattractor} and \ref{sm:fig:henonterms}), and an independent recomputation
of the maximal Lyapunov exponent gives $\lambda_1\simeq0.01062$, consistent with the value
$+0.010657$ of Table~\ref{sm:tab:pairs}. Along this attractor the cross-term is bounded by
$-1.96\times10^{-5}\lesssim\eta\,xy\lesssim5.10\times10^{-4}$, with root-mean-square value
${\rm RMS}(\eta\,xy)\simeq3.21\times10^{-4}$, to be compared with the two other contributions to
the $y$ update, ${\rm RMS}\big((\sqrt{b}/\kappa)x\big)\simeq2.25\times10^{-1}$ and
${\rm RMS}(c\,y^2)\simeq2.63\times10^{-2}$: the direct contribution of the cross-term to one
iteration is two to three orders of magnitude below the linear and quadratic terms. At fixed
state, the change of one $y$ update produced by the parameter difference
$\Delta\eta=-3\times10^{-4}$ between the two members of Table~\ref{sm:tab:pairs} is
$\Delta y=(\Delta\eta)\,xy$, hence $|\Delta y|\lesssim7.3\times10^{-5}$ over the observed range of
$xy$. The transition therefore does not require a large one-step displacement; it requires the
persistent action of the term on the ordered sequence of Jacobians, together with proximity to a
stability boundary, which the continuation below locates.

\subsection{Continuation of the periodic branches}

The bifurcation structure was resolved by continuation rather than read off a scan. For a
period-$p$ point $z$ of the map $F_\eta$ of Eq.~\eqref{sm:eq:henondef}, the period-doubling
bifurcation at which that orbit loses stability and the doubled orbit is born satisfies
\begin{equation}
  F_\eta^{\,p}(z)=z ,
  \qquad
  \det\!\big(DF_\eta^{\,p}(z)+I\big)=0 ,
  \label{sm:eq:perioddoubling}
\end{equation}
with $F_\eta^{\,p}$ the $p$-fold composition, $DF_\eta^{\,p}$ its Jacobian (the monodromy matrix
of the orbit) and $I$ the identity; the second condition is equivalent to one Floquet multiplier
reaching $-1$. Solving Eq.~\eqref{sm:eq:perioddoubling} for successive periods gives the cascade of
Table~\ref{sm:tab:henoncascade}: as $\eta$ decreases the stable orbit doubles from period 2 to 4,
8, 16 and on to 1024 within the resolution used.

The first member of the cascade is not itself a period-doubling of a period-1 orbit. Continuing
the attracting period-2 branch toward larger $\eta$, its two Floquet multipliers form a
complex-conjugate pair that reaches the unit circle near $\eta\simeq0.38509$, a Neimark-Sacker
instability. The route is therefore a Feigenbaum period-doubling cascade built on a period-2
branch, not the standard cascade that starts from a stable fixed point.

\begin{table}[t]
\caption{Generalised H\'enon map of Eq.~\eqref{sm:eq:henondef} for $(a,b,c,\kappa)=(1.1,0.3,0.4,2)$.
Left: successive period-doubling bifurcations obtained from the continuation conditions of
Eq.~\eqref{sm:eq:perioddoubling}; each row gives the critical cross-term amplitude $\eta$ at which the
orbit of the stated stable period has one Floquet multiplier equal to $-1$ and the doubled orbit
is born. Right: ratios $\delta_n$ of Eq.~\eqref{sm:eq:feigratio} of successive parameter
intervals, formed from the three consecutive doublings listed. The universal value of dissipative
period doubling is $\delta=4.669201609\ldots$}
\label{sm:tab:henoncascade}
\centering
\begin{tabular}[t]{r r r}
\hline\hline
Stable period & New period & Critical $\eta$\\
\hline
2   & 4    & $ 0.214056772808$\\
4   & 8    & $ 0.045323227108$\\
8   & 16   & $ 0.009109757242$\\
16  & 32   & $ 0.000441221752$\\
32  & 64   & $-0.001468403805$\\
64  & 128  & $-0.001880038489$\\
128 & 256  & $-0.001968320562$\\
256 & 512  & $-0.001987233537$\\
512 & 1024 & $-0.001991284376$\\
\hline\hline
\end{tabular}
\hspace{2.5em}
\begin{tabular}[t]{l r}
\hline\hline
Intervals used & Ratio $\delta_n$\\
\hline
$2\to4$, $4\to8$, $8\to16$             & $4.659414$\\
$4\to8$, $8\to16$, $16\to32$           & $4.177576$\\
$8\to16$, $16\to32$, $32\to64$         & $4.539390$\\
$16\to32$, $32\to64$, $64\to128$       & $4.639127$\\
$32\to64$, $64\to128$, $128\to256$     & $4.662721$\\
$64\to128$, $128\to256$, $256\to512$   & $4.667805$\\
$128\to256$, $256\to512$, $512\to1024$ & $4.668903$\\
\hline\hline
\end{tabular}
\end{table}

\subsection{Feigenbaum scaling and accumulation point}

With $\eta_n$ the successive thresholds of Table~\ref{sm:tab:henoncascade} ordered by increasing
period, the ratios of successive parameter intervals,
\begin{equation}
  \delta_n=\frac{\eta_{n-1}-\eta_{n-2}}{\eta_n-\eta_{n-1}} ,
  \label{sm:eq:feigratio}
\end{equation}
approach the universal Feigenbaum constant $\delta=4.669201609\ldots$, and extrapolating the
highest-order thresholds with this scaling gives the accumulation point
\begin{equation}
  \eta_\infty\simeq-0.0019923884 .
  \label{sm:eq:etainf}
\end{equation}
This places the two measured states. The stable-periodic value $\eta=-0.00180$ lies between the
$32\to64$ and the $64\to128$ thresholds, hence in the stable period-64 window of the principal
cascade, in agreement with the measured period; the chaotic value $\eta=-0.00210$ lies beyond the
accumulation point. This is why a cross-term whose direct contribution to one iteration is three
orders of magnitude below the linear terms suffices to turn the periodic orbit chaotic: the
remaining period-doublings before $\eta_\infty$ are compressed into a parameter interval of order
$10^{-4}$.

\begin{figure}[t]
\centering
\includegraphics[width=0.79\textwidth]{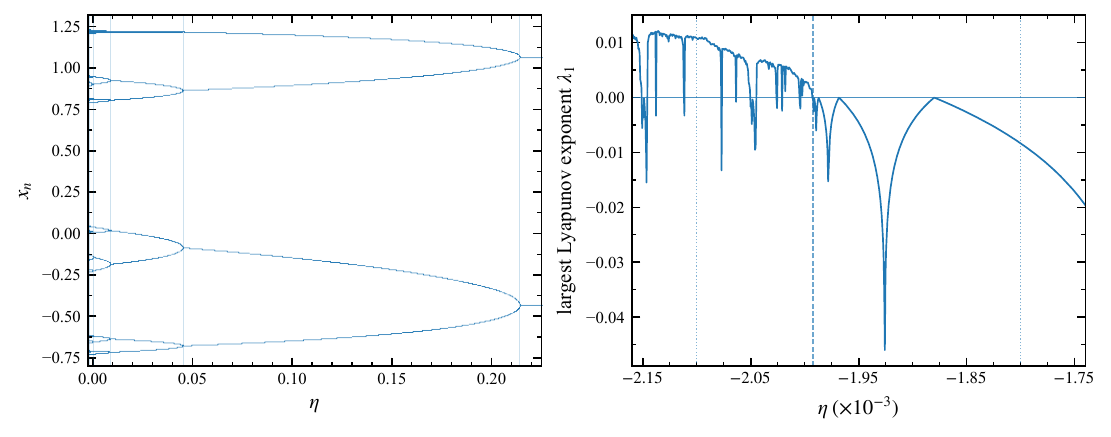}
\caption{Generalised H\'enon map of Eq.~\eqref{sm:eq:henondef} for $(a,b,c,\kappa)=(1.1,0.3,0.4,2)$,
each parameter column computed after a transient of $10^5$ iterations. Left: bifurcation diagram,
attractor coordinate versus the cross-term amplitude $\eta$ on a linear axis. Right: maximal
Lyapunov exponent $\lambda_1$ of Eq.~\eqref{sm:eq:lyap} versus $\eta$ across the high-order
transition region; the accumulation point of the cascade is $\eta_\infty\simeq-0.0019923884$
of Eq.~\eqref{sm:eq:etainf}, and the two values compared in Table~\ref{sm:tab:pairs} are
$\eta=-0.00180$ and $\eta=-0.00210$.}
\label{sm:fig:henonbif}
\end{figure}

\begin{figure}[t]
\centering
\includegraphics[width=0.7\textwidth]{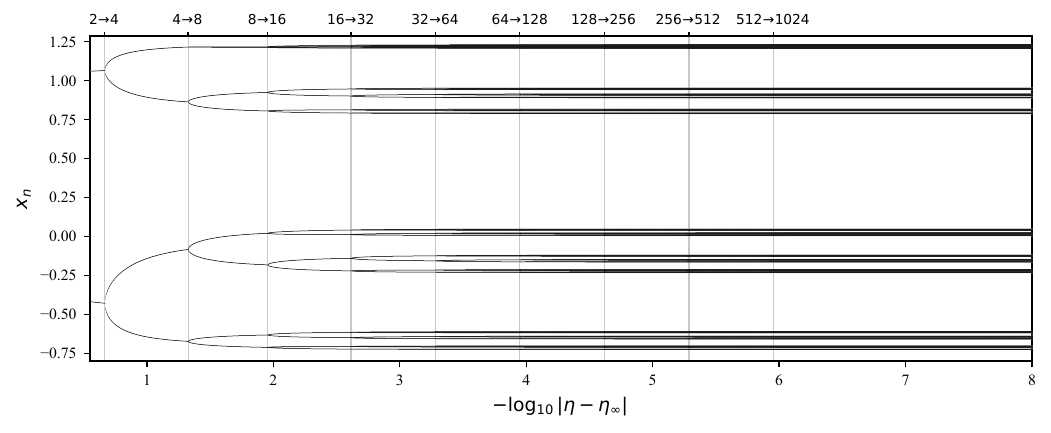}
\caption{Bifurcation diagram of the generalised H\'enon map for $(a,b,c,\kappa)=(1.1,0.3,0.4,2)$,
from the stable period-2 branch onward, plotted against $-\log_{10}|\eta_\infty-\eta|$ with
$\eta_\infty\simeq-0.0019923884$. Guide lines mark the period-doubling bifurcations $2\to4$, $4\to8$,
$8\to16$, $16\to32$, $32\to64$, $64\to128$, $128\to256$, $256\to512$ and $512\to1024$ of
Table~\ref{sm:tab:henoncascade}.}
\label{sm:fig:henonbiflog}
\end{figure}

The maximal Lyapunov exponent changes sign in the same region
(Fig.~\ref{sm:fig:henonbif}, right). Beyond $\eta_\infty$ the chaotic regime is interrupted by
narrow periodic windows, so $\lambda_1$ is not positive at every parameter value above the
accumulation point: no single critical $\eta$ separates the periodic and chaotic regimes globally, which is why
the comparisons of Section~\ref{sm:sec:henon} are made between adjacent bracketed states rather
than across a coarse sweep. Using $-\log_{10}|\eta_\infty-\eta|$ as horizontal coordinate
converts the geometric contraction of successive intervals by the factor $\delta$ into a nearly
uniform spacing (Fig.~\ref{sm:fig:henonbiflog}).

\section{Period-doubling route in the Ikeda map}
\label{sm:sec:ikedaroute}

\subsection{Map and scan}

The Ikeda map of a ring cavity containing a Kerr medium is used in rescaled unit-injection form,
\begin{equation}
  z_{n+1}=1+B\,z_n\exp\!\big[i\big(\phi+\beta|z_n|^2\big)\big] ,
  \qquad z_n=x_n+iy_n ,
  \label{sm:eq:ikeda}
\end{equation}
with $B\in(0,1)$ the round-trip amplitude transmission (so $1-B$ is the loss), $\phi$ the linear
phase and $\beta$ the Kerr coefficient. Writing $\bm{r}_n=(x_n,y_n)^{\top}$,
$\theta_n=\phi+\beta\lVert\bm{r}_n\rVert^2$, $R(\theta)$ the rotation matrix of angle $\theta$ and
$Q=R(\pi/2)$,
\begin{equation}
  \bm{r}_{n+1}=\begin{pmatrix}1\\0\end{pmatrix}+B\,R(\theta_n)\,\bm{r}_n ,
  \qquad
  \J_n=B\,R(\theta_n)\big(I+2\beta\,Q\,\bm{r}_n\bm{r}_n^{\top}\big) ,
  \label{sm:eq:ikedajac}
\end{equation}
so every Jacobian is the product of a scalar contraction $B$, an orthogonal rotation $R(\theta_n)$
and a state-dependent non-orthogonal shear $I+2\beta Q\bm{r}_n\bm{r}_n^{\top}$. Since $\det R=1$
and $\det(I+2\beta Q\bm{r}\bm{r}^{\top})=1$,
\begin{equation}
  \det\J_n=B^2
  \label{sm:eq:detB2}
\end{equation}
exactly, at every point of every orbit: changing $\beta$ changes the rotations and the shears
without changing the local area contraction. One consequence deserves note: wherever the
eigenvalues of $\J_n$ form a complex-conjugate pair, $\rmax(\J_n)=\sqrt{\det\J_n}=B$
exactly, independent of the state. The distribution of $\ln\rmax$ therefore has an
atom at $\ln B$, common to both members of every fixed-loss pair; it contributes nothing
to the distances between the two states and produces the plateau visible in the
cumulative distributions.

\begin{figure}[t]
\centering
\includegraphics[width=0.79\textwidth]{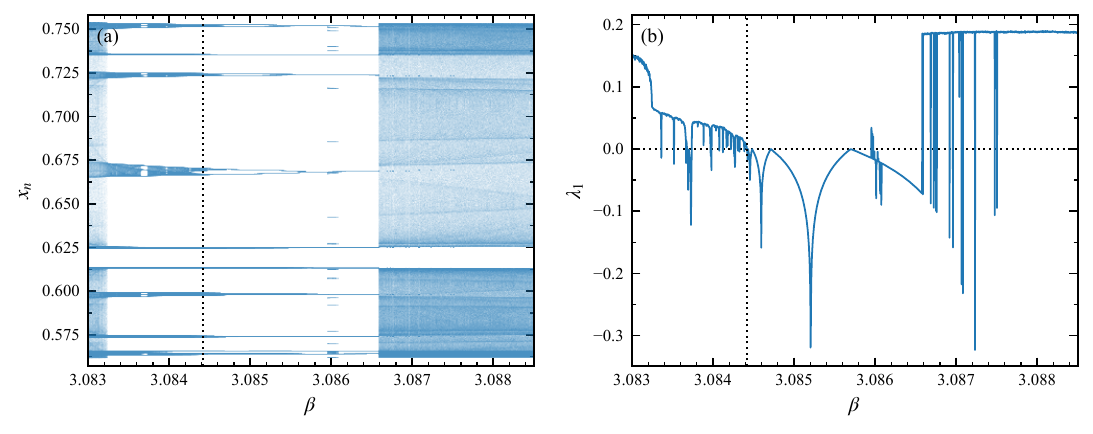}
\caption{Ikeda map of Eq.~\eqref{sm:eq:ikeda} at round-trip transmission $B=0.65$ and linear
phase $\phi=1.5$, each parameter column computed after a $10^5$-iterate transient. Left:
bifurcation diagram, attractor coordinate versus the Kerr coefficient $\beta$. Right: maximal
Lyapunov exponent $\lambda_1$ of Eq.~\eqref{sm:eq:lyap} over the same interval. Markers indicate
the saddle-node creation of the primitive period-10 branch for $\beta_{\rm SN}=3.0876934290$, the
period doublings of Table~\ref{sm:tab:ikedacascade}, the accumulation point
$\beta_\infty\simeq3.0844182475$, and the two values $\beta=3.0846$ and $\beta=3.0836$ of
Table~\ref{sm:tab:pairs}.}
\label{sm:fig:ikedabif}
\end{figure}

This section uses $B=0.65$, $\phi=1.5$, for which the pair of Table~\ref{sm:tab:pairs} is
$\beta=3.0846$ (stable periodic orbit) and $\beta=3.0836$ (chaotic). Their ratio of one-step shift
to exponent shift is $|\Delta h_1/\Delta\lambda_1|=0.07\%$, with
$\Delta h_1=(8.42\pm2.83)\times10^{-5}$ from $32$ contiguous blocks of $3.2\times10^{4}$
iterations per member, measured as the mean of the per-block differences, which need not
equal the difference of the two rounded means in the last digit;
for the pair with $B=0.55$, $\phi=0.5$ the ratio is $0.04\%$, with
$\Delta h_1=-4.29\times10^{-5}$ and $95\%$ confidence interval
$[-4.75,-3.83]\times10^{-5}$, and there $h_1$ moves in the direction opposite to
$\lambda_1$; neither interval contains zero. Both stable
periodic orbits have $h_1>0>\lambda_1$. A direct scan in $\beta$ (Fig.~\ref{sm:fig:ikedabif})
shows a high-period branch followed by a dense sequence of period doublings and then chaos, the
stable periodic orbit realized as a period-$40$ orbit.

\subsection{Cascade and Feigenbaum scaling}

The periodic branch was continued directly, solving for a period-$p$ point $z$ of the map
$F_\beta$ of Eq.~\eqref{sm:eq:ikeda} the same pair of conditions as in Eq.~\eqref{sm:eq:perioddoubling},
$F_\beta^{\,p}(z)=z$ and $\det[DF_\beta^{\,p}(z)+I]=0$. The resulting cascade is
$10\to20\to40\to80\to160\to320\to640\to\cdots$, with the thresholds of
Table~\ref{sm:tab:ikedacascade}. With $\Delta_n=\beta_{p_n}-\beta_{p_{n+1}}$ the successive
interval lengths, the ratios $\Delta_{n-1}/\Delta_n$ are $4.2501$, $4.5564$, $4.6468$ and
$4.6642$, converging toward $\delta=4.669201609\ldots$, and the corresponding accumulation point
is
\begin{equation}
  \beta_\infty\simeq3.0844182475 .
  \label{sm:eq:betainf}
\end{equation}
The base period of this Feigenbaum cascade is ten rather than one. Since
$3.084714303443>3.0846>3.084482935788$, the value $\beta=3.0846$ lies between the $20\to40$ and
the $40\to80$ thresholds and has a stable period-40 orbit, as measured; its distance to the
accumulation point is only $3.0846-\beta_\infty\simeq1.82\times10^{-4}$, so a change of order
$10^{-3}$ in $\beta$ crosses all remaining doublings, and $\beta=3.0836$ lies beyond
$\beta_\infty$.

\begin{table}[t]
\caption{Successive period-doubling bifurcations of the primitive period-10 branch of the Ikeda
map of Eq.~\eqref{sm:eq:ikeda} for $B=0.65$ and $\phi=1.5$, obtained by continuation. Each column
gives the critical Kerr coefficient $\beta_p$ at which the orbit of stable period $p$ has one
Floquet multiplier equal to $-1$ and the orbit of period $2p$ is born.}
\label{sm:tab:ikedacascade}
\centering
\begin{tabular}{l c c c c c c}
\hline\hline
Stable period $p$  & 10 & 20 & 40 & 80 & 160 & 320\\
New period $2p$    & 20 & 40 & 80 & 160 & 320 & 640\\
Critical $\beta_p$ & $3.0856976295$ & $3.0847143034$ & $3.0844829358$
                   & $3.0844321567$ & $3.0844212289$ & $3.0844188860$\\
\hline\hline
\end{tabular}
\end{table}

\subsection{Origin of the period-10 branch}

\begin{figure}[t]
\centering
\includegraphics[width=0.79\textwidth]{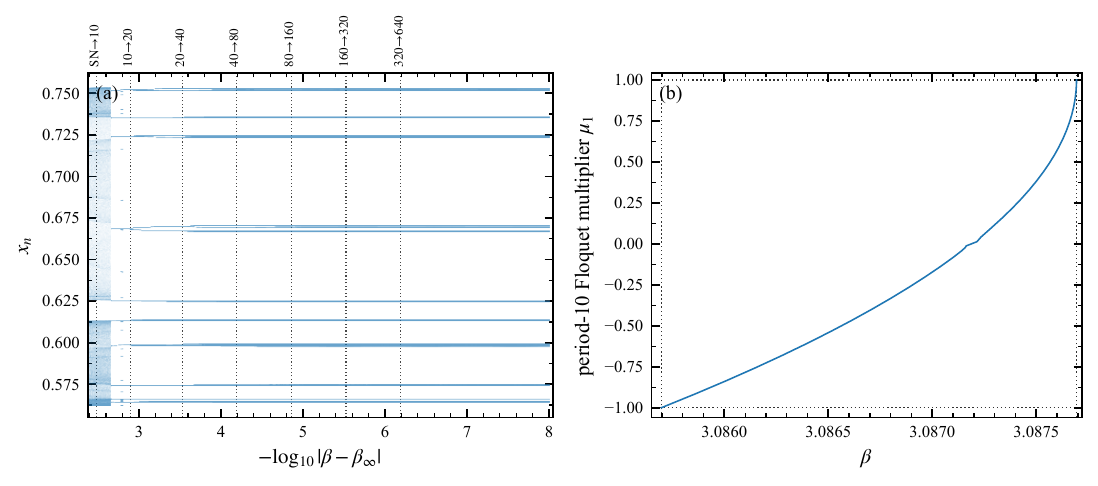}
\caption{Ikeda map for $B=0.65$ and $\phi=1.5$. Left: bifurcation diagram plotted against
$-\log_{10}|\beta-\beta_\infty|$ with $\beta_\infty\simeq3.0844182475$
of Eq.~\eqref{sm:eq:betainf}, extending from the saddle-node birth of the primitive period-10
branch for $\beta_{\rm SN}=3.0876934290$ through the sequence $10\to20\to40\to\cdots$ of
Table~\ref{sm:tab:ikedacascade}. Right: dominant Floquet multiplier $\mu_1$ of the primitive
period-10 orbit, that is the larger eigenvalue modulus of the monodromy matrix $DF_\beta^{\,10}$,
versus $\beta$; the branch terminates at $\beta_{\rm SN}$ where $\mu_1=+1$ and reaches
$\mu_1=-1$ for $\beta=3.0856976295$.}
\label{sm:fig:ikedalog}
\end{figure}

The period-10 orbit is not inherited from a period-5 doubling. Continuing the primitive period-10
solution toward larger $\beta$ and following its Floquet multipliers $\mu_1,\mu_2$, that is the
eigenvalues of the monodromy matrix $DF_\beta^{\,10}$, the branch terminates at
\begin{equation}
  \beta_{\rm SN}=3.0876934290 ,\qquad \mu_1=+1 ,
  \label{sm:eq:betasn}
\end{equation}
the signature of a saddle-node (tangent) bifurcation creating a stable and an unstable primitive
period-10 orbit as a pair. At that point, the orbit is genuinely of minimal period ten: it
satisfies neither the period-1, nor the period-2, nor the period-5 return condition. The second
multiplier is fixed by the determinant identity \eqref{sm:eq:detB2}, since
$\mu_1\mu_2=\det DF^{\,10}=B^{20}=1.81245458\times10^{-4}$, so $\mu_2\simeq1.81\times10^{-4}$ at
the saddle-node. As $\beta$ decreases along the stable period-10 branch, the dominant multiplier
moves continuously from $+1$ to $-1$, reaching $-1$ for $\beta=3.0856976295$, which produces the
first doubling $10\to20$ (Fig.~\ref{sm:fig:ikedalog}, right). The local sequence is therefore:
saddle-node birth of period ten, then $10\to20\to40\to80\to\cdots$, then Feigenbaum accumulation
and chaos.

Two limitations are explicit. First, a saddle-node at the edge of a periodic window is, in
one-dimensional maps, the local mechanism associated with type-I Pomeau-Manneville intermittency
on the neighbouring chaotic side; here the map is two-dimensional and the scan shows
multistability, so these calculations establish the saddle-node origin of the branch, not an
intermittency scaling law for an observed trajectory, which would require a separate analysis of
laminar episodes near $\beta_{\rm SN}$. Second, the analogous statement for the pair with $B=0.55$ and $\phi=0.5$ rests on the
period-doubling sequence of Section~\ref{sm:subsec:b055cascade} rather than on a
continuation of its branch.

\subsection{The period-doubling sequence for \texorpdfstring{$B=0.55$}{B = 0.55},
\texorpdfstring{$\phi=0.5$}{phi = 0.5}}
\label{sm:subsec:b055cascade}

The pair with $B=0.55$ and $\phi=0.5$ of Table~\ref{sm:tab:pairs} sits on a period-doubling
cascade. The approximate doubling locations are $\beta\simeq2.04025$ ($8\to16$),
$2.043825$ ($16\to32$), $2.044597$ ($32\to64$), $2.044763$ ($64\to128$) and $2.044799$
($128\to256$). The successive interval ratios are approximately $4.63$, $4.65$ and $4.61$,
approaching the universal value $\delta=4.669201609\ldots$ from below at this resolution,
and the accumulation point is estimated as $\beta_\infty\simeq2.04481$. The stable
periodic member of the pair, $\beta=2.0442$, therefore lies on the stable period-32 branch
of that cascade, in agreement with the measured period, and the chaotic member,
$\beta=2.0452$, lies beyond the accumulation point. These are
locations read from a scan and refined, not continuation solutions of
Eq.~\eqref{sm:eq:perioddoubling}, so the quoted ratios are indicative of Feigenbaum scaling
and are not a measurement of $\delta$.

\subsection{Loss-driven contrast}

To separate a periodic-to-chaotic transition produced at fixed loss, by reordering the tangent
maps, from one produced by changing the loss itself, the transmission itself was moved, from
$B=0.6420$ to $B=0.6422$, for fixed linear phase and Kerr coefficient
$(\phi,\beta)=(1.5,3.0836)$. The exponent changes sign,
$\lambda_1=-0.022524$ to $+0.022982$, and the one-step rate moves together with it:
$|\Delta h_1/\Delta\lambda_1|=6\%$, two orders of magnitude above the fixed-loss pairs of
Table~\ref{sm:tab:pairs}. The change $\Delta(2\ln B)=6\times10^{-4}$ of the pinned exponent sum
$\lambda_1+\lambda_2=2\ln B$
accounts for only a quarter of the measured one-step shift $\Delta h_1=0.0027$, so that shift is
not itself the determinant of the transition; the comparison contrasts two ways of crossing the
periodic-chaotic boundary, not an additional mechanism.

\section{The crisis transition at fixed loss}
\label{sm:sec:crisis}

\subsection{A second transition, distinct from the cascade}

\begin{figure}[t]
\centering
\includegraphics[width=0.79\textwidth]{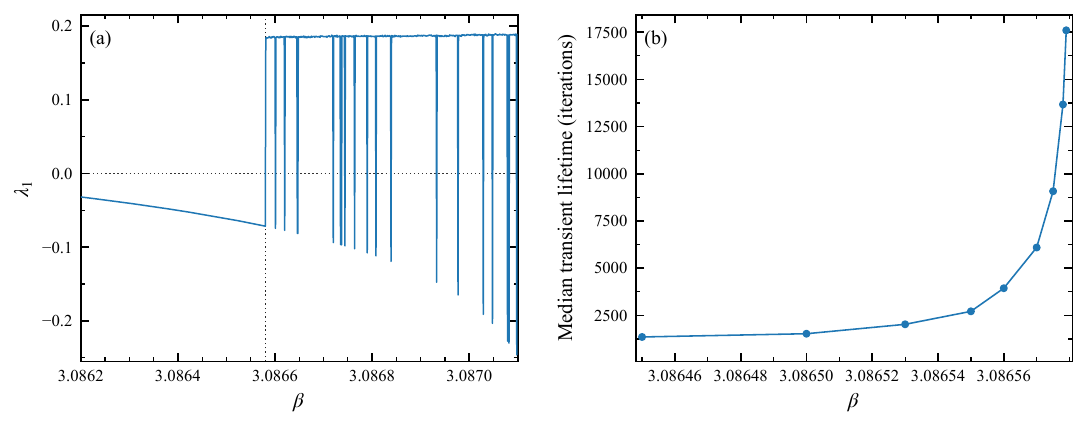}
\caption{Crisis region of the Ikeda map of Eq.~\eqref{sm:eq:ikeda} for $B=0.65$ and $\phi=1.5$.
Left: maximal Lyapunov exponent $\lambda_1$ of Eq.~\eqref{sm:eq:lyap} versus the Kerr coefficient
$\beta$ in a fine scan from a fixed initial state, each value after a $10^5$-iterate transient;
the dashed line marks $\beta_c\simeq3.08658$. Right: median lifetime of chaotic transients as a function of the Kerr coefficient
$\beta$ below the crisis, $\beta<\beta_c$. The horizontal-axis labels use the offset
$+3.086$, so, for example, $0.00050$ corresponds to $\beta=3.08650$.
The lifetime is the number of iterations before capture by the stable period-$10$
attractor, for initial conditions sampled from the chaotic attractor at $\beta=3.0878$.}
\label{sm:fig:crisisscan}
\end{figure}

For the same $B=0.65$ and $\phi=1.5$, a second stable-periodic-to-chaotic transition occurs well above the
accumulation point $\beta_\infty\simeq3.0844182475$. A fine scan in $\beta$ from a fixed initial
state shows an abrupt change from a stable period-10 orbit to strong chaos near
$\beta_c\simeq3.08658$, with $\lambda_1(3.086580)\simeq-0.071$ and
$\lambda_1(3.086581)\simeq+0.186$, that is a change of about $0.26$ across a parameter step of
$10^{-6}$ (Fig.~\ref{sm:fig:crisisscan}, left). The isolated negative spikes visible above the
transition are narrow periodic windows embedded in the chaotic regime.

Continuation of the period-10 branch shows that it survives beyond $\beta_c$ and terminates only
at the saddle-node point $\beta_{\rm SN}=3.0876934290$ of Eq.~\eqref{sm:eq:betasn}, where
$\mu_1=+1$ and $\mu_2\simeq1.81\times10^{-4}$. The transition at $\beta_c$ therefore cannot be a
local bifurcation in which the stable period-10 orbit disappears: over part of the interval
$\beta_c<\beta<\beta_{\rm SN}$, the stable period-10 attractor and a chaotic invariant set
coexist, and which one is reached depends on the initial condition.

To test the crisis interpretation, initial conditions were sampled from the chaotic attractor for
$\beta=3.0878$ and evolved at values of $\beta$ below $\beta_c$. They display long chaotic
transients before capture by the stable period-10 attractor, with a median lifetime growing
rapidly as $\beta\to\beta_c^{-}$ (Fig.~\ref{sm:fig:crisisscan}, right). Together with the
survival of the periodic branch, this supports the picture: for $\beta<\beta_c$ a stable period-10
attractor coexists with a chaotic saddle, for $\beta>\beta_c$ a chaotic attractor is accessible,
that is a boundary-crisis-type transition. The qualifier is deliberate: the evidence is a
diverging transient lifetime plus a surviving periodic branch, and no global continuation of the
invariant manifolds was performed, which is what identifying the collision itself would require.

\subsection{One-step statistics across the crisis}

The two sides were sampled at $\beta_R=3.08657$ (stable period 10) and $\beta_C=3.08660$
(chaotic), with measured exponents $\lambda_1^R\simeq-0.0701$ and $\lambda_1^C\simeq+0.1846$. The
one-step statistics of Table~\ref{sm:tab:crisis} remain close: a change of order $10^{-3}$ to
$10^{-2}$ in the one-step diagnostics accompanies a change of about $0.25$ in the asymptotic
exponent, a one-step shift equal to $2.4\%$ of that jump. The two jump values quoted in this
section refer to two different brackets: about $0.26$ across the fine step
$\beta=3.086580$ to $3.086581$ of the scan above, and about $0.25$ across the wider
bracket $\beta_R$ to $\beta_C$ on which the one-step statistics are sampled here. Here the
two invariant sets are
geometrically different, so the near-coincidence of their one-step statistics cannot be ascribed
to continuity of the invariant measure across a window boundary.

\begin{figure}[t]
\centering
\includegraphics[width=0.7\textwidth]{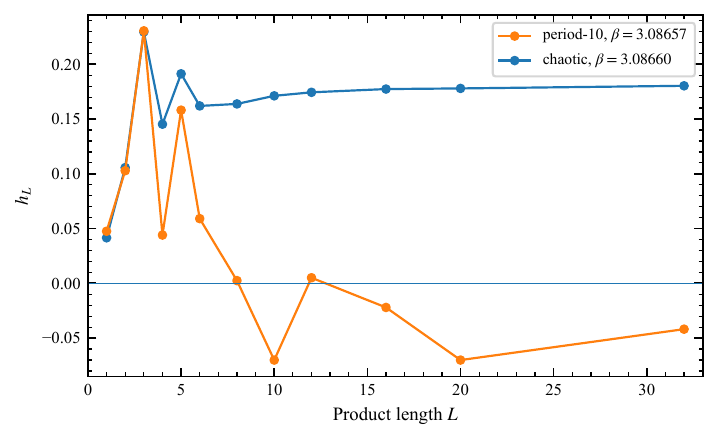}
\caption{Ordered-product rate $h_L$ of Eq.~\eqref{sm:eq:hL} versus product length $L$ on the two
sides of the crisis of the Ikeda map for $B=0.65$ and $\phi=1.5$: stable period-10 state at
$\beta_R=3.08657$ and chaotic state at $\beta_C=3.08660$, from $1.024\times10^6$ retained
Jacobians after a $10^5$-iterate transient on the chaotic side and exact orbit averages on the
periodic side.}
\label{sm:fig:crisishL}
\end{figure}

Writing the singular value decomposition $\J_n=\Umat_n\Smat_n\Vmat_n^{\top}$, with $\Smat_n$
diagonal and $\Umat_n,\Vmat_n$ orthogonal (for these non-normal Jacobians
$\smax>\rmax$, so the frames $\Umat_n\neq\Vmat_n$ contain information absent from any
spectrum), two consecutive steps give
\begin{equation}
  \J_{n+1}\J_n=\Umat_{n+1}\Smat_{n+1}\,\Cmat_n\,\Smat_n\Vmat_n^{\top} ,
  \qquad
  \Cmat_n=\Vmat_{n+1}^{\top}\Umat_n ,
  \label{sm:eq:sigmac}
\end{equation}
so a long product is governed by the alternating chain of local gains $\Smat_n$, which one-step
statistics measure, and inter-step frame couplings $\Cmat_n$, which they do not. A scalar summary
of $\Cmat_n$, the mean angle between consecutive dominant singular directions, moves only from
$0.651$ to $0.661$ rad across the crisis (Table~\ref{sm:tab:crisis}): the transition does not
follow from a change of mean rotation angle either. What separates the two sides is the ordered
product itself, $h_{10}=-0.0701$ against $+0.1712$ (Fig.~\ref{sm:fig:crisishL}).

\begin{table}[t]
\caption{One-step and ordered-product diagnostics on the two sides of the crisis of the Ikeda map
for $B=0.65$ and $\phi=1.5$. $\rmax(\J_n)$ is the one-step spectral radius, $\smax(\J_n)$ the
one-step largest singular value, $\avg{\cdot}$ the orbit average, $h_{10}$ the ordered-product
rate of Eq.~\eqref{sm:eq:hL} at $L=10$, $\lambda_1$ the maximal Lyapunov exponent of
Eq.~\eqref{sm:eq:lyap}, and the angle is the mean angle between the dominant singular directions
of consecutive Jacobians, the scalar summary used in the crisis analysis. Sampling follows the
production protocol of Section~\ref{sm:sec:methods}: $1.024\times10^6$ retained Jacobians after a
$10^5$-iterate transient on the chaotic side, exact orbit averages over the ten starting phases on
the periodic side, the same sampling as Fig.~\ref{sm:fig:crisishL}. The chaotic row uses the
$32$-window baseline of Section~\ref{sm:sec:crisis}, the single value used for this state
throughout.}
\label{sm:tab:crisis}
\centering
\begin{tabular}{l c c c c c c}
\hline\hline
Kerr coefficient & $\avg{\ln\rmax(\J_n)}$ & $\avg{\rmax(\J_n)}$ & $\avg{\ln\smax(\J_n)}$
 & Angle (rad) & $h_{10}$ & $\lambda_1$\\
\hline
$\beta_R=3.08657$ (period 10) & $0.0476$ & $1.1217$ & $0.7588$ & $0.651$ & $-0.0701$ & $-0.0701$\\
$\beta_C=3.08660$ (chaotic)   & $0.0416$ & $1.1173$ & $0.7643$ & $0.661$ & $+0.1712$ & $+0.1846$\\
\hline\hline
\end{tabular}
\end{table}

\subsection{Same-parameter coexistence}

The comparison above brackets the crisis with two adjacent values of $\beta$, one per side. A
direct test holds $\beta$ fixed and varies only the initial condition. For $\beta=3.08660$, the
single value used for the chaotic side above, the initial condition $(x_0,y_0)=(0,0.8)$ converges
to the stable primitive period-10 orbit with $\lambda_1=-0.074241$, while $(x_0,y_0)=(0.1,0.1)$
converges to a chaotic orbit with $\lambda_1=+0.184584$: both invariant sets are reached at one and
the same parameter value, direct evidence of coexistence rather than an inference from a narrow
bracket. On a $61\times61$ grid of initial conditions, $101$ of the $3721$ points converge to the
period-10 orbit and the remainder to the chaotic attractor. The one-step and ordered-product
diagnostics repeat the pattern of Table~\ref{sm:tab:crisis}: the mean one-step rate changes from
$\avg{\ln\rmax(\J_n)}=0.047451$ on the periodic side to $0.041583$ on the chaotic side, while the
ten-step ordered rate changes from $h_{10}=-0.074241$ to $+0.171200$; on the periodic side
$h_{10}=\lambda_1$ exactly, by the period-10 exactness identity of Section~\ref{sm:sec:methods}.
For the chaotic state, the baseline values are the means over $32$ non-overlapping windows of
$1.024\times10^6$ iterations, $\lambda_1=0.184637\pm0.000072$,
$h_1=0.041583\pm0.000022$ and
$h_{10}=0.171200\pm0.000053$, with window standard deviations $2.1\times10^{-4}$,
$6.3\times10^{-5}$ and
$1.5\times10^{-4}$. The first production window alone gives
$h_1=0.041735$ and $h_{10}=0.171224$ and the single controlled-baseline trajectory quoted above gives
$\lambda_1=+0.184584$.  The $32$-window means are the values used in the Letter. From these six-digit values the one-step rate moves by
$\Delta h_1=0.047451-0.041583=0.005868$ against
$\Delta\lambda_1=0.184637-(-0.074241)=0.258878$, that is
$|\Delta h_1/\Delta\lambda_1|=2.3\%$, the ratio quoted in the Letter.

The two coexisting attractors also differ in dimension. For the chaotic attractor,
$\lambda_1=0.184637$ and the exact pairing $\lambda_1+\lambda_2=2\ln B$ give
$\lambda_2=-1.046203$, hence a Kaplan-Yorke dimension
$D_{\rm KY}=1+\lambda_1/|\lambda_2|=1.176$, the fractal transverse structure visible in the
nested magnifications of Fig.~3 of the Letter; the periodic orbit is ten points.

\begin{figure}[t]
\centering
\includegraphics[width=0.6\textwidth]{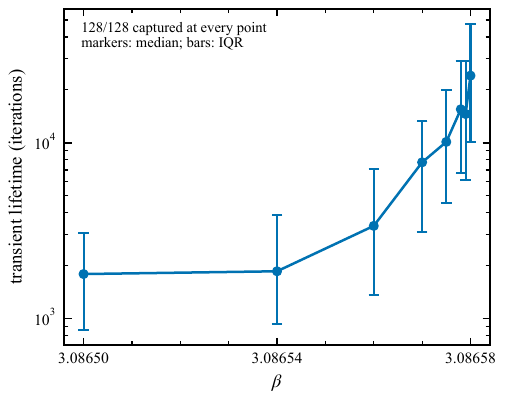}
\caption{Chaotic-transient lifetime, in iterations before capture by the stable period-10 orbit,
versus the Kerr coefficient $\beta$ over $[3.08650,3.08658]$, for $128$ initial states sampled at
each value of $\beta$; all $128$ are captured at every plotted $\beta$ (in-panel text). Points
show the median lifetime; vertical bars show the interquartile range.}
\label{sm:fig:crisislifetimes}
\end{figure}

To test the crisis interpretation independently of a single sampled trajectory, chaotic-transient
lifetimes were measured for $\beta\in[3.08650,3.08658]$, below $\beta_c$, with $128$ initial states
sampled at each value of $\beta$, each followed until capture by the stable period-10 orbit
(Fig.~\ref{sm:fig:crisislifetimes}); this sampling protocol applies to the lifetime measurement
only, and not to Table~\ref{sm:tab:crisis} above.

\section{Reordering tests}
\label{sm:sec:ordering}

Two families of tests act on the temporal arrangement of the stored Jacobians while leaving
their one-step marginal law untouched. For the Ikeda map, the construction is exact: every
Jacobian on the attractor has determinant $B^2$ by Eq.~\eqref{sm:eq:detB2}, so any permutation of
the stored sequence preserves both the one-step marginals and the area-contraction constraint.

\emph{Full reordering.} The stored Jacobians of one state are resampled without regard to their
original order, and the growth rate of the reordered product is compared with the ordered one.
This is not a neutral reference: random frames sample the transiently amplifying directions
efficiently, so the amplification available in the local gains $\Smat_n$ of
Eq.~\eqref{sm:eq:sigmac} is realized rather than cancelled, and the reordered products are
expanding for both members of a pair, periodic and chaotic alike. The same mechanism produces
heavy-tailed amplification in random multiplicative processes with non-orthogonal
eigenvectors~\cite{troudeKesten1,troudeKesten2}, which is why the informative
comparison is between the two endogenous orderings and not between an endogenous ordering and a
random one.

\emph{Exchanging contiguous segments of the tangent sequence.} Contiguous segments of fixed
length $m$ are kept intact and only the segments are exchanged, so the order is destroyed above
the scale $m$ and preserved below it. Sweeping $m$ interpolates between the fully reordered limit
and the true dynamics. Over $1.024\times10^6$ stored Jacobians and $32$ permutations per segment
length, the rate of the stable periodic member of a pair first returns to a negative value, its
chaotic partner remaining positive, at the segment lengths and rates of
Table~\ref{sm:tab:blockexchange}.

These measured thresholds, $m^*=8$, $16$ and $4$, are of the same order as the separation
lengths $L^*=8$, $10$ and $7$ of Table~\ref{sm:tab:pairs}, and they are not equal to them. The two
lengths measure different things: the separation length is the number of consecutive Jacobians
whose ordered product separates the two true sequences, whereas the segment threshold is the
amount of consecutive order that must be preserved for that separation to survive a rearrangement
of the segments, an operation that also replaces one coupling at every junction. The thresholds
are case dependent.

For the generalised H\'enon pair, the two rates at $m^*=8$ have nearly equal magnitudes and
opposite signs. The near equality of the two magnitudes is accidental: the check was
repeated with $32$ independent permutations of the $1.024\times10^6$ stored Jacobians,
returning $-0.01086\pm0.00003$ for the stable periodic member and
$+0.01086\pm0.00001$ for its chaotic partner, the two rates agreeing in magnitude to
$8\times10^{-7}$, far inside either confidence interval; the two magnitudes are therefore
equal to the resolution of the measurement, and the opposite signs are reproduced by an
independent computation rather than by a sign error in a transcription. The two Ikeda pairs of the same table have magnitudes that
differ by a factor of nearly two, so no symmetry between the two members is implied.

\begin{table}[t]
\caption{Growth rates of the rearranged tangent sequences as a function of the preserved
segment length $m$, for the three periodic-chaotic pairs of Table~\ref{sm:tab:pairs}: ensemble
means over $32$ independent segment permutations of $1.024\times10^{6}$ stored Jacobians,
quoted to four decimals. The $95\%$ confidence half-widths lie between $5\times10^{-10}$ and
$2.3\times10^{-4}$ and are largest for the shortest segments; at the bold entries, the smallest
$m^*$ for which the two rates have opposite signs with both confidence intervals excluding
zero, they are $-0.010860\pm0.000031$ and $+0.010861\pm0.000013$ for the generalised
H\'enon pair, $-0.028911\pm0.000102$ and $+0.050454\pm0.000064$ for $B=0.65$, and
$-0.043169\pm0.000060$ and $+0.027531\pm0.000027$ for $B=0.55$. For $m$ beyond $m^*$ each
rate approaches the Lyapunov exponent of its own unpermuted sequence.}
\label{sm:tab:blockexchange}
\centering
\small
\setlength{\tabcolsep}{4.5pt}
\begin{tabular}{c cc cc cc}
\hline\hline
& \multicolumn{2}{c}{Generalised H\'enon} & \multicolumn{2}{c}{Ikeda, $B=0.65$}
& \multicolumn{2}{c}{Ikeda, $B=0.55$}\\
$m$ & Periodic & Chaotic & Periodic & Chaotic & Periodic & Chaotic\\
\hline
1 & $+0.2138$ & $+0.2137$ & $+0.1191$ & $+0.1193$ & $+0.1150$ & $+0.1156$\\
2 & $+0.0294$ & $+0.0298$ & $+0.1284$ & $+0.0152$ & $+0.0092$ & $+0.0155$\\
4 & $+0.0121$ & $+0.0158$ & $+0.0757$ & $+0.0571$ & $\mathbf{-0.0432}$ & $\mathbf{+0.0275}$\\
8 & $\mathbf{-0.0109}$ & $\mathbf{+0.0109}$ & $+0.0034$ & $+0.0539$ & $-0.0834$ & $+0.0272$\\
16 & $-0.0083$ & $+0.0107$ & $\mathbf{-0.0289}$ & $\mathbf{+0.0505}$ & $-0.0840$ & $+0.0270$\\
32 & $-0.0083$ & $+0.0107$ & $-0.0545$ & $+0.0489$ & $-0.0831$ & $+0.0270$\\
64 & $-0.0083$ & $+0.0107$ & $-0.0674$ & $+0.0481$ & $-0.0831$ & $+0.0270$\\
\hline\hline
\end{tabular}
\end{table}

\section{Feedback on the tangent ordering}
\label{sm:sec:control}

\subsection{Modulated map and control rule}

The actuated quantity is the round-trip phase of the cavity map,
\begin{equation}
  z_{n+1}=1+B\,z_n\exp\!\big[i\big(\phi+\beta|z_n|^2+u_n\big)\big] ,
  \qquad
  u_n\in\{-\varepsilon,0,+\varepsilon\} ,
  \qquad
  \varepsilon=10^{-4}\ \text{rad} ,
  \label{sm:eq:modmap}
\end{equation}
so $|u_n|\le\varepsilon$ at every step. Because $u_n$ enters only the phase, the identity
$\det\J_n=B^2$ of Eq.~\eqref{sm:eq:detB2} still holds exactly: the modulation reorders rotations
and shears at strictly fixed area contraction, that is at fixed cavity loss.

The action is chosen at every step by minimizing the forecast tangent growth over the next $H$
steps, recomputed anew at each iteration. The horizon $H$ is the number of steps over which a
candidate action is evaluated; it is kept distinct from the composition length $L$ of
Eq.~\eqref{sm:eq:hL} even where the two share the same numerical value. Every run reported here
uses $H=10$, the composition length at which $h_L$ resolves the stable periodic orbit from the
chaotic orbit for the parameter values used.
For each candidate action $u$, an $H$-step forecast ordered product $\widehat{\bm{P}}^{(H)}_{n,u}$ is built
from the known map and the current state, with $u$ applied at the current step only and $u=0$ at
the following $H-1$ forecast steps; the forecast rate and the applied action are
\begin{equation}
  G_n(u)=\frac{1}{H}\ln\big\lVert\widehat{\bm{P}}^{(H)}_{n,u}\,\qv_n\big\rVert ,
  \qquad
  u_n=\arg\min_{u\in\{-\varepsilon,0,+\varepsilon\}}G_n(u) .
  \label{sm:eq:rule}
\end{equation}
$G_n(u)$ is the forecast criterion, and the rule applies the action that minimizes it.
The tangent vector is then propagated
with the Jacobian of the action actually selected. The rule is model-based; it locates, continues
and targets no orbit, and uses no information beyond the current state and the known map.

Three per-iteration exponents characterize the closed loop and are reported separately, never
aggregated. $\lambda_{\rm traj}$ is the tangent growth rate of Eq.~\eqref{sm:eq:lyap} along the
realized controlled trajectory with the realized action sequence held fixed; because
Eq.~\eqref{sm:eq:rule} contains a discrete minimization, it is not automatically the derivative
exponent of the full feedback law. $\lambda_{\rm sep}$ is the separation rate of independently
controlled nearby pairs, two trajectories started a distance $d_0$ apart that each recompute
their own action at every iteration, fitted to the logarithm of their separation and reported as
the median over an ensemble of $160$ pairs, built from five initial separations
$d_0=10^{-10},\dots,10^{-6}$, eight perturbation directions and four phases of the controlled
orbit, each run over $1024$ iterations. $\lambda_{\rm fb}$ is, for the orbit-targeting benchmark
below, the logarithm of the dominant eigenvalue modulus of the closed-loop return matrix,
including the linear response of the feedback action to the perturbed state. Two cost measures
accompany them: $u_{\rm rms}=\avg{u_n^2}^{1/2}$ and the active fraction $f_{\rm act}$, the
fraction of iterations with $u_n\neq0$.

\subsection{The parameter values used for the control, and where the rule has no effect}

For $B=0.65$, $\phi=1.5$, $\beta=3.0836$, the chaotic state of Table~\ref{sm:tab:pairs}, the rule
reverses the sign of the realized exponent, $\lambda_{\rm traj}=+0.047112\to-0.004568$,
identically for all $16$ initial conditions sampled across the uncontrolled attractor (ensemble
values differing from the long-run values of Table~\ref{sm:tab:pairs} in the sixth decimal). The
reversal is not paid out of the local budget: the mean one-step spectral rate increases, from
$h_1=+0.047660$ to $+0.049864$, while $\avg{\ln\smax(\J_n)}$ changes by only
$-5.7\times10^{-5}$. What reverses is the ordered product, $h_{10}=+0.042279$ without control
against $-0.004568$ with control. The stricter test agrees: all $160$ independently controlled
nearby pairs converge, with median $\lambda_{\rm sep}=-0.004568$, and no pair ever selects a
different action from its partner.

Two comparisons fix what this establishes. A \emph{constant bias} $u_n=2\times10^{-5}$, within
the same bound and amounting to a static shift of the linear phase $\phi$, also stabilizes the
map for $\beta=3.0836$, on a different orbit of period $30$, with
$\lambda_{\rm traj}=-0.035941$; the two rules separate under variation of the nonlinearity, which
is what the parameter grids below measure. \emph{Open-loop replay} repeats the period-10 action
sequence onto which the closed loop settles, applied without recomputing the action from the current
state. Across the $25$-value benchmark grid of Kerr coefficients, it stabilizes all $16$ initial
conditions at $2$ points, against $23$ for the feedback rule, so what survives detuning is the
state dependence of the rule and not the waveform it generates.

\subsection{Comparison with orbit-targeting control}

The benchmark is the classical scheme that stabilizes a preselected unstable periodic orbit by
small parameter kicks~\cite{ogy,shinbrot1993,boccaletti,roy1992}, implemented on the same map, through the same phase parameter, under the
same bound, with the same initial conditions and run length, its target orbit selected before any
closed-loop performance was examined. A search through period 12 returns one qualifying primitive
period-10 unstable orbit, with return residual $2.81\times10^{-13}$ and Floquet multipliers
approximately $-1.8803$ and $-9.64\times10^{-5}$; the search is not claimed to be exhaustive.

Performance is reported on five separate axes (Table~\ref{sm:tab:control}). No single number ranks
the methods: the axes measure different things, are not commensurable, and do not define an
economic cost. For $\beta=3.0836$, the value at which it was designed, the orbit-targeting
scheme is the more economical,
acting only inside a trigger neighbourhood of its target orbit, on $9.7\%$ of the iterations, with
a root-mean-square phase action more than five orders of magnitude below that of the tangent rule,
which acts at every step. The last column measures a design requirement: the tangent rule is
fixed once and applied unchanged across the grid, whereas the orbit-targeting scheme must have
its target orbit relocated and its gains recomputed at each new parameter value, a relocation for
which tracking schemes exist~\cite{gills1992}. The penalized variant adds a quadratic cost $\gamma u^2$ with
$\gamma=10^{5}$ to $G_n(u)$ before the minimization; it gives $\lambda_{\rm traj}=-0.102412$,
negative for all $16$ initial conditions and so entered as $16/16$ in the first column of
Table~\ref{sm:tab:control}, while the stricter independently controlled nearby-trajectory test,
in which both members of a pair recompute their own action at every iteration, succeeds for $13$
of those $16$ initial conditions.

\begin{table}[t]
\caption{Control comparison for $B=0.65$, $\phi=1.5$. All methods act through the phase $u_n$ of
Eq.~\eqref{sm:eq:modmap} under the same bound $|u_n|\le\varepsilon=10^{-4}$ rad. The success column
counts the initial conditions, out of $16$ sampled across the uncontrolled attractor for
$\beta=3.0836$, for which the realized exponent $\lambda_{\rm traj}$ is negative within the common
run length; $u_{\rm rms}=\avg{u_n^2}^{1/2}$; $f_{\rm act}$ is the fraction of iterations with
$u_n\neq0$; the stabilization time is the mean number of iterations before capture by the
stabilized state; the last column counts the values of the Kerr coefficient $\beta$, out of $25$
evenly spaced from $3.0830$ to $3.0842$, at which all $16$ initial conditions are stabilized with
the design fixed for $\beta=3.0836$, except in the last row where the design is
redone at every $\beta$. The current detuning
result is the $53$-value grid of Section~\ref{sm:subsec:mismatch}. Stabilization times and actuation statistics are operational,
method-specific
quantities, each defined by the validated criterion of its own method, and are not a universal
physical cost. Dashes mark quantities not measured for that method.}
\label{sm:tab:control}
\centering
\begin{tabular}{l c c c c c}
\hline\hline
Method & Success & $u_{\rm rms}$ & $f_{\rm act}$ & Stabilization time & $\beta$ grid\\
\hline
Ten-step tangent rule, Eq.~\eqref{sm:eq:rule} & $16/16$ & $1.000\times10^{-4}$ & $1.000$ & $2978$ & $23/25$\\
Penalized tangent rule                        & $16/16$ & $8.744\times10^{-5}$ & $0.765$ & $507$  & $23/25$\\
Constant bias $u_n=2\times10^{-5}$            & $16/16$ & $2.000\times10^{-5}$ & $1.000$ & $708$  & $3/25$\\
Open-loop replay                              & --      & --                   & --      & --     & $2/25$\\
Orbit targeting, design at $\beta=3.0836$    & $16/16$ & $4.314\times10^{-10}$& $0.097$ & $15560$& $1/25$\\
Orbit targeting, redesigned at every $\beta$  & --      & --                   & --      & --     & $11/25$\\
\hline\hline
\end{tabular}
\end{table}

The two control schemes stabilize by different mechanisms. Along the stabilized target orbit of the
orbit-targeting scheme, the tangent product of the uncontrolled map remains expanding: with the
realized action sequence frozen, $\lambda_{\rm traj}=+0.063143$. This is not a failure of that
scheme, since its stability includes the change of the feedback action produced by the perturbed
state, which a frozen sequence excludes by construction; including it gives
$\lambda_{\rm fb}=-0.924709$, and its independently controlled separation rate, the analogue of
$\lambda_{\rm sep}$ for that scheme, is $-0.804315$. The tangent rule instead makes the realized
tangent product itself contracting.

\begin{figure}[t]
\centering
\includegraphics[width=\textwidth]{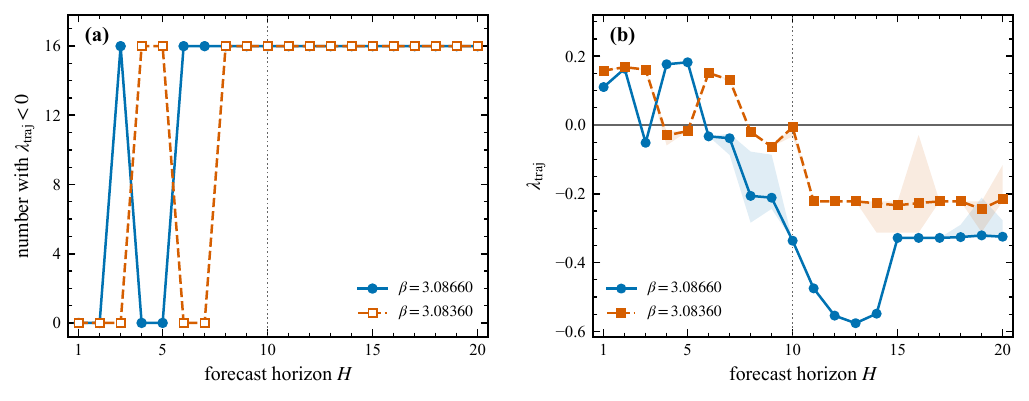}
\caption{Dependence of the controlled exponent on the controller's forecast horizon $H$ of
Eq.~\eqref{sm:eq:rule}, for every integer $H=1,\ldots,20$ and the same $16$ initial conditions as
Table~\ref{sm:tab:control}, for the crisis parameter value $\beta=3.08660$ and for
$\beta=3.08360$. Each run follows $10^5$ controlled settling iterations by
$5\times10^5$ evaluation iterations. Left: number of the $16$ initial conditions with negative
$\lambda_{\rm traj}$ versus $H$, for both values of $\beta$. Right: $\lambda_{\rm traj}$ versus
$H$, with the range over the $16$ initial conditions shaded, for both values of $\beta$; both
curves reach $16/16$ negative by $H=10$.}
\label{sm:fig:controlhorizonH1H20}
\end{figure}

The horizon $H=10$ was chosen as the composition length at which $h_L$ resolves the stable
periodic orbit from the chaotic orbit for $\beta=3.0836$, the same length that resolves
the same-parameter coexistence pair of Section~\ref{sm:sec:crisis} ($\beta=3.08660$,
Table~\ref{sm:tab:pairs}), and was not tuned for performance. The dependence of the
stabilization on the horizon $H$ of Eq.~\eqref{sm:eq:rule} was then measured directly.
Holding the phase bound and every other controller setting fixed, the same $16$
initial conditions were run for every integer $H=1,\ldots,20$ for $\beta=3.08660$, each with
$10^5$ controlled settling iterations followed by $5\times10^5$ evaluation iterations: not all
$16$ controlled exponents are negative for $H=1$, $2$, $4$ and $5$, while all $16$ are negative for
$H=3$ and for every $H=6,\ldots,20$ (Fig.~\ref{sm:fig:controlhorizonH1H20}). The dependence on $H$
is therefore nonmonotonic, and $H=10$ is retained because $L=10$ was selected independently from
the uncontrolled ordered-product comparison, not because $H=10$ gives the most negative exponent.
The corresponding horizon dependence for $\beta=3.08360$ is shown in
Fig.~\ref{sm:fig:controlhorizonH1H20}; both parameter values reach $16/16$ negative controlled
exponents by $H=10$.

\subsection{The same control rule in the strong chaos above the crisis}

\begin{figure}[t]
\centering
\includegraphics[width=0.78\textwidth]{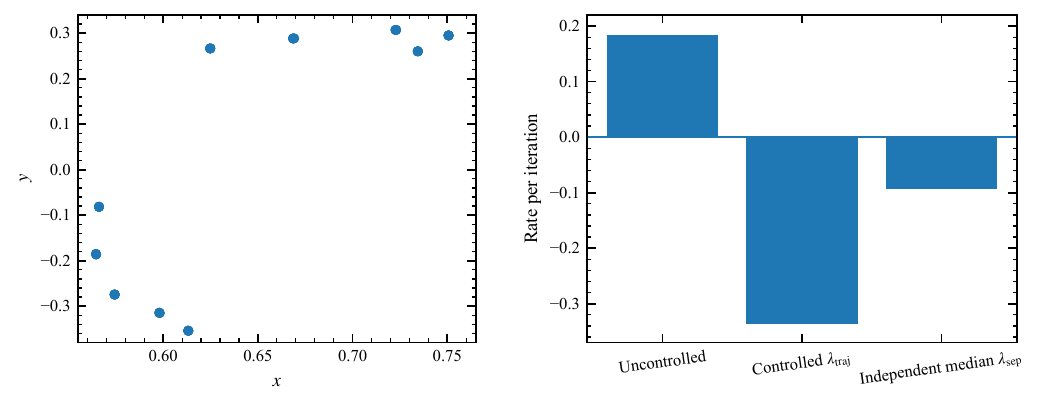}
\caption{Ikeda map for $B=0.65$, $\phi=1.5$, Kerr coefficient $\beta=3.0866$, controlled by the
rule of Eq.~\eqref{sm:eq:rule} with horizon $H=10$ and bound $\varepsilon=10^{-4}$ rad. Left:
orbit of period $70$ reached after capture due to control, ${\rm Im}\,z_n$ versus ${\rm Re}\,z_n$.
Right: per-iteration exponents, namely the realized-trajectory exponent $\lambda_{\rm traj}$ of
the uncontrolled and of the controlled dynamics and the independently controlled separation rate
$\lambda_{\rm sep}$ over the $160$ nearby pairs defined by five initial separations $10^{-10}$ to
$10^{-6}$, eight perturbation directions and four phases of the controlled orbit, each pair run
over $1024$ iterations with both members recomputing their own action.}
\label{sm:fig:controlorbit}
\end{figure}

The control rule was applied unchanged for $\beta=3.0866$, above the crisis of
Section~\ref{sm:sec:crisis}, where the uncontrolled exponent is
$\lambda_1^{\rm unc}\simeq+0.1846$,
four times the value for $\beta=3.0836$. For $16$ different initial conditions evolved with the same
parameters and controller, sampled across the uncontrolled chaotic attractor, the controlled
trajectory exponent falls in $\lambda_{\rm traj}\in[-0.336301,-0.336279]$, mean $-0.336291$,
the feedback settling onto a controlled orbit of
period $70$ (Fig.~\ref{sm:fig:controlorbit}, left), with $u_{\rm rms}\simeq9.71\times10^{-5}$ and
$f_{\rm act}\simeq0.943$, that is with the full allowed amplitude used on most iterations. A
broader basin test on a $7\times7$ grid of $49$ initial conditions covering $[-2,2]^2$ gives
negative $\lambda_{\rm traj}$ in all $49$ runs, within
$-0.33657\lesssim\lambda_{\rm traj}\lesssim-0.33606$: broad practical attraction to the controlled
periodic regime, not a proof of global stability.

\begin{figure}[t]
\centering
\includegraphics[width=0.7\textwidth]{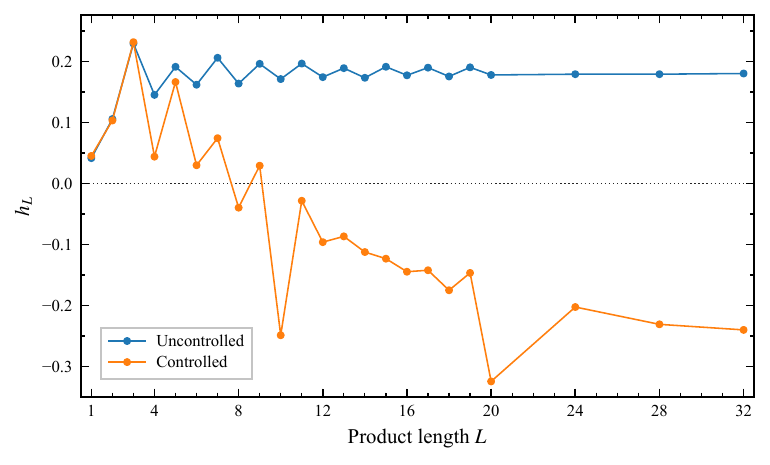}
\caption{Ordered-product rate $h_L$ of Eq.~\eqref{sm:eq:hL} versus product length $L$ for the
uncontrolled and the controlled dynamics of the Ikeda map for $B=0.65$, $\phi=1.5$,
$\beta=3.0866$, the control being the rule of Eq.~\eqref{sm:eq:rule} with horizon $H=10$ and
bound $\varepsilon=10^{-4}$ rad.}
\label{sm:fig:controlhL}
\end{figure}

As for the parameter value used above, the stabilization does not act by suppressing local
stretching. The one-step diagnostics barely move, $\avg{\ln\rmax(\J_n)}$ from $0.04158$ to
$0.04517$,
that is
upward, and $\avg{\ln\smax(\J_n)}$ from $0.76432$ to $0.75896$, while the ordered products reverse
sign: $h_{10}$ from $+0.1712$ to $-0.2489$ and $h_{20}$ from $+0.1780$ to $-0.3246$
(Fig.~\ref{sm:fig:controlhL}).

The $160$-pair test was repeated for this parameter value. The median fitted separation rate is
$\lambda_{\rm sep}\simeq-0.09355$, using the explicit fitting rule defined above; $141$ of
$160$ fitted rates are negative;
$140$ of $160$ final
separations fall below $10^{-10}$; and
$136$ of $160$ pairs select identical actions throughout
(Fig.~\ref{sm:fig:controlorbit}, right). The remaining pairs occasionally cross a switching
surface of the discrete minimization in Eq.~\eqref{sm:eq:rule} and then select different actions,
after which they are no longer driven by the same tangent sequence; this differs from the case
$\beta=3.0836$, where all $160$ pairs converged and no pair ever disagreed. These numbers support strong
practical stabilization of a strongly chaotic state with a bounded $10^{-4}$ phase action,
together with an explicit limitation: along its realized action sequence, the controlled orbit is
strongly contracting, but a closed-loop law built on a discrete minimization is neither uniformly
smooth nor uniformly contracting in the stricter nearby-pair sense.

\subsection{Robustness to a mismatch between the controller's model and the actual map}
\label{sm:subsec:mismatch}

\begin{figure}[t]
\centering
\includegraphics[width=0.6\textwidth]{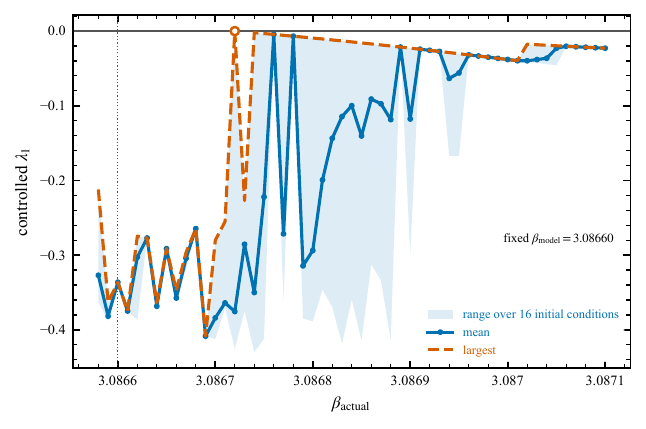}
\caption{Controlled Lyapunov exponent of the Ikeda map for $B=0.65$, $\phi=1.5$, when the
controller of Eq.~\eqref{sm:eq:rule} forecasts with $\beta_{\rm model}=3.08660$ while the map is
evaluated at $53$ actual values $\beta_{\rm actual}$ spanning $[3.08658,3.08710]$ in steps of
$10^{-5}$, $16$ initial conditions at each value. Every exponent is negative; the largest is
$-4.92\times10^{-5}$, for $\beta_{\rm actual}=3.08672$, where a run of $3.2\times10^{6}$
iterations gives $-5.56\times10^{-5}$ with $95\%$ confidence interval
$[-5.93,-5.20]\times10^{-5}$.}
\label{sm:fig:mismatch}
\end{figure}

The purpose of this test is to ask whether a controller built at one parameter value remains
effective when the actual map parameter is slightly different, without retuning the controller. The controller of Eq.~\eqref{sm:eq:rule} is constructed with
$\beta_{\rm model}=3.08660$, so every internal forecast uses this value, while the map it acts on
is evaluated at $53$ actual values $\beta_{\rm actual}$ spanning $[3.08658,3.08710]$ in steps of
$10^{-5}$, with $16$ initial conditions at each value, that is $848$ controlled trajectories, and
without retuning the controller at any of them. For every tested actual value, all $16$ initial
conditions give a negative controlled exponent; the largest exponent over the full tested grid is
$-4.92\times10^{-5}$, attained for $\beta_{\rm actual}=3.08672$; at that value of
$\beta_{\rm actual}$ a longer run, $3.2\times10^{6}$ iterations for each of the $16$ initial
conditions in $32$ blocks of $10^{5}$, gives $\lambda_1=-5.56\times10^{-5}$ with $95\%$
confidence interval $[-5.93,-5.20]\times10^{-5}$ and block-to-block standard deviation
$1.0\times10^{-5}$, every one of the $32$ blocks being negative.
The mean controlled exponent
at $\beta_{\rm actual}=\beta_{\rm model}$ is $-0.336295$ (Fig.~\ref{sm:fig:mismatch}). 
The result is established on the tested interval of
$\beta_{\rm actual}$ and on the $16$ initial conditions used at each of its values.

\subsection{Scope of the control results}

These calculations support the following. The bounded phase rule \eqref{sm:eq:rule}
reverses the sign of the realized tangent growth rate for the two values of the Kerr coefficient
at which it was run, $\beta=3.0836$ and $\beta=3.08660$, at strictly fixed area contraction,
without reducing the mean one-step spectral rate or the non-normal amplification available at each
step, and without locating, continuing or targeting any orbit. With its design fixed once, it
keeps every controlled exponent negative across the $53$-value grid of
Section~\ref{sm:subsec:mismatch}, $848$ controlled trajectories in all. A constant phase bias
within the same bound stabilizes the map for $\beta=3.0836$, on an orbit of period $30$; what the
rule of Eq.~\eqref{sm:eq:rule} adds is that one fixed design covers the whole detuning grid
without locating any orbit.
A coarser $25$-value grid of Kerr coefficients from $3.0830$ to $3.0842$, on which the rule
stabilizes all $16$ initial conditions at $23$ points, is reported in
Table~\ref{sm:tab:control} for the comparison between control schemes. The
comparison with orbit-targeting control measures robustness to detuning without redesign, on
axes reported separately, and is not a statement that one method is superior.

\end{document}